\documentclass[a4paper,11pt]{article}

\usepackage{amsmath,amsthm,amsbsy,amssymb,amsfonts,graphicx,ascmac,mathrsfs,bm,accents,cite,xcolor,color}

\makeatletter
\catcode`\@=11
\@addtoreset{equation}{section}

\makeatother

\renewcommand{\thefootnote}{\arabic{footnote}}

\newcommand{\Exp}[1]{\operatorname{e}^{#1}}
\newcommand{\rmd}{{\mathrm{d}}}
\newcommand{\nn}{\nonumber}
\newcommand{\Lie}{\pounds}
\newcommand{\gLie}{\hat{\pounds}}

\newcommand{\cF}{\mathcal F}
\newcommand{\cG}{\mathcal G}
\newcommand{\cH}{\mathcal H}
\newcommand{\cL}{\mathcal L}
\newcommand{\cM}{\mathcal M}
\newcommand{\cN}{\mathcal N}
\newcommand{\cO}{\mathcal O}

\newcommand{\cR}{\mathcal R}

\newcommand{\cU}{\mathcal U}
\newcommand{\cV}{\mathcal V}

\newcommand{\bB}{\mathbf{B}}
\newcommand{\bC}{\mathbf{C}}

\newcommand{\sfv}{\mathsf v}

\newcommand{\sfV}{\mathsf V}

\newcommand{\ours}{\mathcal{S}}

\newcommand{\fv}{{\tilde{\mathfrak{v}}}}

\newcommand{\lR}{{\mathbb R}}
\newcommand{\GL}{\mathrm{GL}}
\newcommand{\SL}{\mathrm{SL}}
\newcommand{\OO}{\mathrm{O}}
\begin{document}

\begin{titlepage}
\renewcommand{\thefootnote}{\fnsymbol{footnote}}


\vspace*{1.0cm}

\begin{center}
{\textbf{\Large Finite Transformations \\[-1.5mm]
 in \\
Doubled and Exceptional Space
}}%
\end{center}
\vspace{1.0cm}

\centerline{
\textsc{Soo-Jong Rey} $^{a,b}$%
\footnote{E-mail address: sjrey@snu.ac.kr} \hskip0.5cm  
\textsc{Yuho Sakatani} $^{a, b} $%
\footnote{E-mail address: sakatani@snu.ac.kr}%
}

\vspace{0.6cm}

\begin{center}
${}^a${\it School of Physics \& Astronomy and Center for Theoretical Physics \\
Seoul National University, Seoul 08862 \rm KOREA}\\
\vskip0.2cm
${}^b${\it B.W. Lee Center for Fields, Gravity \& Strings \\
Institute for Basic Sciences, Daejeon 34047 \rm KOREA}
\end{center}

\vspace*{1cm}

\begin{abstract}
In the double field theory, gauge symmetries are realized as generalized diffeomorphisms in the doubled spacetime. By consistency of the theory, dependence of tensor fields on the doubled coordinates is strongly constrained. This causes finite transformation law highly complicated, both technically and conceptually. In this paper, we propose a new, physically intuitive approach to finite gauge transformations by utilizing untwisted form of vector fields. In our approach, finite gauge transformation law is expressed in terms of diffeomorphisms on a maximal null subspace on which dynamical degrees of freedom live and of a local rotation of this null subspace embedded inside the doubled space. We show that our finite transformation automatically satisfies the composition law. We also show that ours is free from the so-called Papadopoulos problem, so can describe background with nontrivial three-form flux $H_3$. Added advantage of our approach is straightforward applicability to general extended field theories. We demonstrate this by explicitly obtaining finite transformation law in the SL(5) exceptional field theory. 
\end{abstract}

\thispagestyle{empty}
\end{titlepage}

\setcounter{footnote}{0}

\tableofcontents

\section{Introduction}

It is well-known that a string compactified on a $d$-torus, $\mathbb{T}^d = (\mathbb{S}^1)^d$ possess $\OO(d,d, \mathbb{Z})$ $T$-duality symmetry, exchanging winding and center-of-mass momentum modes. 
Actually the $\OO(d,d)$ duality rotation transforms not only these zero-modes but also string oscillator modes among themselves. 
In particular, the duality rotation acts non-trivially on the massless states of the graviton, Kalb-Ramond field and dilaton as well as Ramond-Ramond $p$-forms. 
To put the $T$-duality symmetry manifest in low-energy effective field theory of these massless states, a novel approach called the double field theory (DFT) was developed \cite{Siegel:1993th,Siegel:1993xq,Hull:2009mi,Hull:2009zb,Hohm:2010jy,Hohm:2010pp}. 

In DFT, the dimensions of spacetime are doubled and all massless fields are covariantly extended over the doubled spacetime. 
Yet, the extension should maintain the same degrees of freedom. 
This requirement puts a consistency condition that the extended fields need to be covariantly constrained. 
This so-called strong constraint puts actual physical excitations to depend only on half of the doubled spacetime coordinates. 
This constraint renders the structure of doubled spacetime quite different from that of the conventional spacetime. 
As such, the geometry of the doubled spacetime -- both local and global aspects -- was not fully understood to date. 

In addition to the global $\OO(d,d)$ duality symmetry, the DFT is also invariant under local gauge transformations, which is the generalization of the $d$-dimensional coordinate transformations or diffeomorphisms and the gauge transformations of Kalb-Ramond $B$-field. 
This combined gauge transformation is called the generalized coordinate transformation \cite{Hohm:2012gk}. 
Better understanding of properties of this generalized coordinate transformation is an important step toward deeper understanding of geometric aspects of DFT. 
Under infinitesimal transformations, the properties are well understood; the generalized coordinate transformations are generated by the generalized Lie derivatives. 
The purpose of this paper is to investigate finite version of the generalized coordinate transformations of doubled spacetime, viz.~the finite gauge transformations of DFT. 

Aspects of the finite generalized coordinate transformations were studied previously. 
After the original work by Hohm and Zwiebach \cite{Hohm:2012gk}, finite generalized coordinate transformations in the doubled spacetime and global aspects of the doubled spacetime were examined from various perspectives \cite{Hohm:2012mf,Park:2013mpa,Lee:2013hma,Hohm:2013bwa,Berman:2014jba,Cederwall:2014kxa,Papadopoulos:2014mxa,Cederwall:2014opa,Papadopoulos:2014ifa,Naseer:2015tia}. 
The finite transformation law is specified by a transformation matrix acting on a (generalized) tensor field. 
In the original approach of Hohm and Zwiebach, the transformation matrix is manifestly $\OO(d,d)$ covariant and has a simple form, yet left several outstanding issues unanswered. 
An issue raised by Papadopoulos \cite{Papadopoulos:2014mxa} is that imposing a patching condition with the transformation matrix puts the three-form field strength $H_3$ of Kalb-Ramond two-form potential an exact form. 
As such, in backgrounds with non-trivial $H_3$-flux, it is not possible to patch the doubled spacetime with their transformation matrix. 
Another issue is that their transformation matrix does not satisfy the composition law in the usual manner \cite{Hohm:2012gk,Hohm:2013bwa,Berman:2014jba}. 
In order to address these issues further, a completely different approach, which uses the untwisted form of vector fields, was proposed by Hull \cite{Hull:2014mxa}. 
In this approach, the transformation matrix was expressed in terms of both a coordinate transformation in a $d$-dimensional null subspace $\cN$ inside the doubled spacetime and a gauge parameter, $\fv_m(x)$, associated with a finite transformation of Kalb-Ramond potential $B_2$. 
If we regard the transformation matrix as a transition function between tensor fields in different patches, it should satisfy the cocycle condition on a triple overlap of patches (which is necessary for the composition law to be satisfied). 
It was found (see section 6.5 of \cite{Hull:2014mxa}) that the transformation matrix indeed satisfies the cocycle condition of composition laws, thus ameliorating the issue raised by Papadopoulos. 

In this paper, we propose a new, physically intuitive approach to finite transformations in DFT, which is manifestly free from the above two issues regarding the finite generalized coordinate transformations. 
Our proposal is similar in spirit but different in core physics details to Hull \cite{Hull:2014mxa}, and also has the desirable feature that allows straightforward extension to EFTs. 
In Hull's approach, finite gauge parameter $\fv_m(x)$ was taken of its own, not associated with any geometric quantities in the doubled spacetime, such as a variation in the dual coordinates. 
In our approach, we take a different approach and relate the finite gauge parameter $\fv_m(x)$ to the variation of an antisymmetric two-form $b_{mn}(x)$ (not to be confused with Kalb-Ramond two-form potential $B_2$) which is a `geometric' quantity inherent to the embedding of null subspace $\cN$ inside the doubled spacetime. 
Facilitated by such geometric structures, we show that our proposed approach to finite generalized coordinate transformation achieves the following features:
\begin{itemize}
\item \textsc{the composition laws}, necessary for putting generalized diffeomorphisms to the same category as finite-dimensional Lie groups or of conventional coordinate transformations in general relativity, 
\item \textsc{restriction-free patching condition}, necessary for describing non-trivial $H_3$ flux background and for formulating immune from the Papadopoulos problem \cite{Papadopoulos:2014mxa}. 
\end{itemize}

An added feature of our approach is that it is straightforwardly generalizable to exceptional spacetimes. 
The M-theory is considered to be a nonperturbative unification of all string theories. 
The M-theory compactified on a torus $\mathbb{T}^d$ is then known to possess a larger duality group, the $U$-duality group \cite{Hull:1994ys}. 
Manifestly $U$-duality covariant formulations of the compactified M-theory have been investigated in various works \cite{Hull:2007zu,Pacheco:2008ps,Hillmann:2009ci,Berman:2010is,Berman:2011pe,Berman:2011cg,Berman:2011jh,Coimbra:2011ky,Coimbra:2012af,Berman:2012vc,Berman:2013eva,Park:2013gaj,Cederwall:2013naa,Cederwall:2013oaa,Cederwall:2007je,Aldazabal:2013mya,Aldazabal:2013via,Rosabal:2014rga}. 
In particular, a new approach, called the exceptional field theory (EFT), was developed recently \cite{Hohm:2013vpa,Hohm:2013uia,Godazgar:2014nqa,Hohm:2014fxa,Musaev:2014lna,Hohm:2015xna,Abzalov:2015ega}. 
While finite transformations in DFT have been studied rather extensively, due to the technical complications in EFT, no concrete proposition has been made for finite gauge transformations in EFT (see, however, \cite{Papadopoulos:2014mxa,Berman:2014jba,Naseer:2015tia} for previous discussions related to the finite transformations in EFT). 
In this paper, we extend our approach to the finite transformation in DFT to the SL(5) EFT and obtain rules for the finite gauge transformations, which also manifestly satisfy the composition laws. 

This paper is organized as follows. 
In section \ref{sec:Finite-Transform}, we recall the finite transformations in infinite-dimensional Lie groups and in general relativity. 
In section \ref{sec:finite-DFT}, using the untwisted form of vector fields, we construct finite generalized coordinate transformations in DFT. 
For comparison, we first review previous approaches. 
We then explain our approach and construct finite transformations in geometric backgrounds. 
We explicitly check that this finite transformations obey the composition laws. 
We next construct finite transformations in non-geometric backgrounds. 
In section \ref{sec:examples}, based on the formalism of section \ref{sec:finite-DFT}, we discuss global aspects of codimension-2 backgrounds in DFT: backgrounds produced by defect NS5-brane or exotic $5^2_2$-brane. 
In section \ref{sec:finite-SL5}, we extend these finite transformations to SL(5) EFT. 
Discussions and future directions are given in section \ref{sec:discussion}.\\

\noindent\textbf{Note added}: While finishing this work, we became aware from private communication of a closely related work by N.~Chaemjumrus and C.~Hull. Their preprint, \cite{Chaemjumrus:2015vap}, which appeared on the arXiv almost two months after our preprint was posted, significantly overlaps with our results and also significantly misinterprets ours. 

\section{Finite Transformations}
\label{sec:Finite-Transform}
Before considering finite generalized coordinate transformations, we first review finite transformations in infinite-dimensional Lie groups and in general relativity, both formulated as differential geometry of metric manifold. 

\subsection{Finite transformations in Lie groups}
We first recall the relation between a Lie algebra $g$ and its Lie group $G$ \cite{helgason} with particular attention to infinite-dimensional situation.%
\footnote{Roughly speaking, by an infinite-dimensional manifold, we mean a manifold modeled on an infinite-dimensional, locally convex vector space, just like a finite-dimensional manifold is modeled on Euclidean space.} 

If $G$ is finite-dimensional, it is well-known that, at least locally around an identity element, the Lie group $G$ is completely described by its Lie algebra $g$. 
This correspondence is given by the exponential map, 
\begin{align}
\mbox{Exp}: \qquad X \in g \quad \rightarrow \quad \exp(X) \in G\,. 
\end{align}
In fact, for every $X \in g$, there always exists a unique analytic homomorphism $v(t)$ of $\mathbb{R}$ into $G$ with $\dot{v}(0) = X$. 
So, the exponential map is a `local diffeomorphism' from the Lie algebra $g$ to the Lie group $G$. 

More explicitly, let $X \in g$ and let $\tilde{X}$ the corresponding left-invariant vector field. 
The flow of the field $\tilde{X}$ is a map $y_X: \mathbb{R} \times G \rightarrow G$ such that
\begin{align}
\frac{\rmd}{\rmd t} y_X(t, g) = \tilde{X}(y_X(t, g)) \quad
\qquad \text{for all} \ \ t \ \ \mbox {and} \ \ 
y_X (0, g) = g\,.
\end{align}
The flow $y_X$ is the solution of this first-order, nonlinear ordinary differential equation. 
For finite-dimensional Lie algebra, the solution always exists and is unique. 
In the case the flow subgroup $y_X(\mathbb{I}, t)$ exists for all $X \in g$, the exponential map is defined by the unit-time map $X \rightarrow y_X (\mathbb{I}, 1)$. 

Note that the analytic homomorphism of $\mathbb{R}$ into $G$ used for the exponential map defines a one-parameter subgroup of the Lie group $G$:
\begin{align}
t \quad \rightarrow \quad \phi_t := \exp (t X) \quad \text{for} \quad X \in g\,.
\end{align}
It has the properties that $(\phi_s)^{-1} = \phi_{-s}$, $\phi_0 = \mathbb{I}$, and  
\begin{align}
\phi_t \circ \phi_s = \phi_s \circ \phi_t \phi_{s + t} \qquad \text{for all} \quad  s, t \in \mathbb{R} \ \  \text{and all}  \ \  X \in g\,.
\end{align}

If $G$ is infinite-dimensional, this correspondence is no longer straightforward. 
There could exist Lie groups that do not admit an exponential map. 
Moreover, even if the exponential map exists, it may not be a local diffeomorphism. 
Furthermore, there could exist infinite-dimensional Lie algebras that do not correspond to any Lie group. 
This is in failure of Lie's third theorem, which states that every finite-dimensional Lie algebra is the Lie algebra attached to some finite-dimensional Lie group. 

One may avoid such pathologies by restricting $G$ to the class of Banach Lie groups, viz.~the Lie groups that are locally defined on Banach spaces and behave like finite-dimensional Lie groups. 
In this case, the exponential map always exists and is a local diffeomorphism. 
Such restrictions would already exclude the important case of diffeomorphism groups. 
This explains why infinite-dimensional differential geometry and infinite-dimensional Lie groups are still under active development. 

\subsection{Finite transformations in (semi-)Riemannian geometry}

We next recall some elements of semi-Riemannian geometry (see \cite{Hawking:1973uf,Wald:1984rg} for concise exposition). 
We consider a differentiable Lorentzian manifold $\cM$ with local coordinates $x^m$. 
For a given vector field $v^m(x)$, we can obtain integral curves or one-parameter families of diffeomorphism, $\phi_s:\,\lR\times \cM\to \cM$ with $s\in\lR$, that has the properties
\begin{align}
 \phi_{s+t}=\phi_s\circ \phi_{t}=\phi_{t}\circ\phi_s\,,\quad 
 (\phi_{s})^{-1}=\phi_{-s}\,,\quad \phi_0(p)=p\,,
\end{align}
for all $s,\,t\in\lR$ and for all $p\in\cM$. 
An integral curve $\phi_s(p)$ through a point $p\in \cM$, of coordinates $x_s^m\bigl(x(p)\bigr)\equiv x^m\bigl(\phi_s(p)\bigr)$, is a solution of the differential equation:
\begin{align}
 \frac{\rmd x_s^m(x)}{\rmd s} = v^m\bigl(x_s(x)\bigr) \,.
\label{eq:integral-curve}
\end{align}
Given the diffeomorphism $\phi_s$ and a function $f(x)$, we can define the pullback of $f(x)$ onto the integral curve by
\begin{align}
 \phi_s^*f(x) \equiv f\bigl(x_s(x)\bigr) \,.
\end{align}
Using this, we can define the pushforward, $\phi_{s*}w$ of a vector field $w$, by
\begin{align}
 (\phi_{s*}w)(x_s)\cdot f(x_s) \equiv w(x)\cdot(\phi_s^*f)(x) \,, 
\label{eq:pushforward}
\end{align}
where the dot denotes the directional derivative, $w(x)\cdot f(x) \equiv w^m(x)\,\partial_m f(x)$.
We can then define the pullback $\phi_s^* \alpha$ of a one-form field $\alpha$ by
\begin{align}
 \langle \phi_s^*\alpha, w\rangle(x) 
 \equiv \langle \alpha, \phi_{s*}w\rangle(x_s) \,,
\end{align}
where $\langle \alpha, v\rangle(x) \equiv \alpha_m(x)\,v^m(x)$ is the conventional canonical pairing of a one-form field and a vector field. 
Since $\phi_s$ is a diffeomorphism, it has the inverse map $(\phi_s)^{-1}=\phi_{-s}$. 
So, we can define the pullback of a vector field or the pushforward of a one-form field by
\begin{align}
 \phi_s^*w(x)\equiv \phi_{-s*}w(x) \,,\quad 
 \phi_{s*}\alpha(x)\equiv \phi_{-s}^*\alpha(x)\,.
\end{align}
From these, the pushforward (or also the pullback, $\phi_s^*=\phi_{-s*}$) of an arbitrary $(k,l)$-tensor field, $T^{m_1\cdots m_k}_{n_1\cdots n_l}(x)$, is straightforwardly given by
\begin{align}
 &T(\alpha_{(1)},\dotsc,\alpha_{(k)}\,;\,w_{(1)},\dotsc, w_{(l)}\bigr)(x)
\nn\\
 &\qquad \qquad \equiv \phi_{s*} T\bigl((\phi_{-s}^*\alpha_{(1)}),\dotsc,(\phi_{-s}^*\alpha_{(k)})\,;\,(\phi_{s*} w_{(1)}),\dotsc, (\phi_{s*} w_{(l)})\bigr)(x_s) \,,
\end{align}
where $\alpha_{(1)},\dotsc,\alpha_{(k)}$ are arbitrary one-form fields and $w_{(1)},\dotsc,w_{(l)}$ are arbitrary vector fields. 
The Lie derivative $\Lie_v$ acting on the $(k,l)$-tensor field is then defined by
\begin{align}
 \Lie_v T(x) 
 \equiv \lim_{s\to 0} \frac{\phi_{-s*}T(x)-T(x)}{s} 
 = -\frac{\rmd}{\rmd s}\phi_{s*}T(x)\biggr\rvert_{s=0} 
 = + \frac{\rmd}{\rmd s}\phi_s^* T(x)\biggr\rvert_{s=0} \,. 
\label{eq:Lie-def}
\end{align}

Up to this point, all definitions and properties are general and valid for arbitrary spacetime manifold. 
In particular, they should be applicable to doubled or exceptional spacetimes (though the definitions might be not sufficient), else it would be difficult to endow a geometric interpretation for such spacetimes. 
Moreover, if there is no constraint on the coordinate dependence of dynamical fields or gauge parameters, taking $f(x)=x^m$ in \eqref{eq:pushforward}, we can completely determine the finite transformation law acting on a vector field $w$\,,
\begin{align}
 \bigl(\phi_{s*}w\bigr)^m(x_s)
 = \frac{\partial x_s^m}{\partial x^n}\,w^n(x) \,,
\label{eq:finite-GR}
\end{align}
and extend it to arbitrary tensor fields. From \eqref{eq:finite-GR}, we easily obtain the conventional Lie derivative:
\begin{align}
 (\Lie_v w)^m(x) = v^n(x)\,\partial_n w^m(x) - w^n(x)\,\partial_n v^m(x) \,. 
\end{align}

For diffeomorphisms of steady flow type, the vector field $v^m(x)$ is independent of the parameter $s$. In this case, the integral curve, $x_s^m(x)$, and the pullback of a $(k,l)$-tensor field can formally be expressed as
\begin{align}
 x_s^m(x) = \Exp{s\,v}\cdot\, x^m \qquad \text{and} \qquad 
 \phi_s^* T(x) = \Exp{s\Lie_v}T(x) \,.
\label{eq:s-independent}
\end{align}
For diffeomorphisms of non-steady flow type, the vector field $v^m(x)$ depends on the parameter $s$. 
If so, the integral curves of $v^m(x)$ at a fixed $s$ (i.e.~streamlines) do not coincide with the pathlines, $x_s^m(x)$, which are defined through \eqref{eq:integral-curve} with the $s$-dependent $v^m(x)$. In this case, expressions like \eqref{eq:s-independent} do not hold. 
In DFT, we sometimes consider this kind of $s$-dependent $v^m(x)$. 

To be self-contained, we also recall the active and passive interpretations of the finite transformation law. 
In the above discussion, we adopted the active point of view, but we can also interpret \eqref{eq:finite-GR} or
\begin{align}
 \bigl(\phi_{-s*}w\bigr)^m(x_{-s})
 = \frac{\partial x_{-s}^m}{\partial x^n}\,w^n(x)
 = \bigl(\phi_s^*w\bigr)^m(x_{-s})
 = \Exp{-s\,v}\,\bigl(\Exp{s\Lie_v}w\bigr)^m(x) \,,
\end{align}
as caused by a passive coordinate transformation. 
In the active point of view, $x_{-s}^m\equiv x^m(\phi_{-s}(p))$ were the coordinate values after the active diffeomorphism, $\phi_{-s}$. 
On the other hand, in the passive interpretation, $x'^m\equiv x^m\circ\phi_{-s}$ are interpreted as the new local coordinates while the physical point, $p\in \cM$, is not changed under the transformation. 
In other words, defining the new coordinates $x'^m$ and the transformed components $w'^m(x')\equiv \phi_{-s*}w^m(x')$, the formula
\begin{align}
 w'^m(x') = \frac{\partial x'^m}{\partial x^n}\,w^n(x)
 = \Exp{-s\,v}\,\bigl(\Exp{s\Lie_v}w\bigr)^m(x) \,,
\end{align}
is interpreted as induced by the coordinate transformation, $x^m\to x'^m$. 
This is simply a change in the interpretation. 
Formally, if one compares this with the pullback in the active point of view \eqref{eq:s-independent}, one just changes in the sign in the transformed coordinates, $x'^m=\Exp{-s\,v}x^m$\,.

\subsection{Finite transformations in double / exceptional field theories}
\label{sec:finite-extended}
We now generalize the above considerations to the situation that the manifold $\cM$ is doubled or exceptional spacetime. 
When we defined DFT or EFT, the key difference from the general relativity is the presence of strong constraint. 
The constraint imposes that, if we introduce a suitable local coordinates, $x^M=(x^m,\,\tilde{x}_i)$ ($m=1,\dotsc,d,\ i=1,\dotsc,n$),\footnote{Doubled spacetime corresponds to $n = d$, while exceptional spacetime corresponds to $n \ne d$.} all tensor fields and gauge parameters must be independent of the dual coordinates, $\tilde{x}_i$ (i.e.~$\tilde{\partial}^i =0$). 
Then, the function $f(x)$ in the left-hand side of \eqref{eq:pushforward} cannot depend on the dual coordinates, $\tilde{x}_i$ either. 
In the following, we denote a finitely transformed generalized vector, with a gauge parameter $V^M(x)$ and a parameter $s$, by $W_{(s,V)}^M(x)$. 
As the finite gauge transformation can be identified with the pullback under a finite diffeomorphism, $x^M\to x_s^M$, it should satisfy
\begin{align}
 & \boxed{~
 W_{(s,V)}(x)\cdot f(x) 
 \equiv W^M(x_s)\,\frac{\partial f}{\partial x_s^M}(x(x_s))
 = \frac{\partial x^n}{\partial x_s^m}\,w^m(x_s)\,\partial_n f(x(x_s))\,.~}
\label{eq:WsV-def}
\end{align}
Here, we used the section constraint $\tilde{\partial}^n f(x)=0$ and $(\partial x^n/\partial \tilde{x}^s_m)=0$ in the second equality. 
As the first equality in \eqref{eq:WsV-def} is the defining equation for $W_{(s,V)}^M(x)$, the disappearance of the dual components $(W_{(s,V)})_i(x)$ in the rightmost side means that we cannot determine $(W_{(s,V)})_i(x)$ only from geometric considerations. 
Put differently, in DFT and EFT, there is a room for a modification of the dual components $(W_{(s,V)})_i(x)$ from the conventional transformation law. 

In DFT, where $x^M=(x^m,\,\tilde{x}_m)$ are doubled coordinates and $V^M=(v^m,\,\tilde{v}_m)$, $W^M=(w^m,\,\tilde{w}_m)$ are doubled vector fields, the gauge symmetries are generated by the generalized Lie derivative \cite{Hull:2009mi,Hull:2009zb}:
\begin{align}
 \bigl(\delta_V W\bigr)^M = V^N\,\partial_N W^M - \bigl(\partial_N V^M-\partial^M V_N\bigr)\,W^N 
 \equiv \bigl(\gLie_V W\bigr)^M \, . 
\end{align}
This can be identified with an infinitesimal diffeomorphism in the doubled spacetime $\cM$. 
With the section constraint, $\tilde{\partial}^m =0$ imposed, the generalized Lie derivative of DFT becomes%
\footnote{In this paper, we use the convention for the matrix representation, $W^M=(w^m,\,\tilde{w}_m)$, following \cite{Hull:2014mxa}, although $W^M=(\tilde{w}_m,\,w^m)$ is more commonly used in the literature.}
\begin{align}
\boxed{~
 \gLie_V W^M(x) = \begin{pmatrix}
 (\Lie_v w)^m(x) \cr
 (\Lie_v\tilde{w})_m(x) + (\partial_m\tilde{v}_n -\partial_n\tilde{v}_m)(x)\, w^n(x) 
 \end{pmatrix} \,,~}
\label{eq:gen-Lie-constraint}
\end{align}
where $\Lie_v$ is the conventional Lie derivative along the vector field $v(x)=v^m(x)\,\partial_m$. 
As is clear from \eqref{eq:gen-Lie-constraint}, the dual components (the lower-half components) of $(\gLie_V W)^M(x)$ are indeed modified from the conventional Lie derivative in the doubled spacetime. 

Extension to EFT is straightforward. 
Once the generalized Lie derivative, $\gLie_V W^M(x)$, is prescribed, we first solve the first-order ordinary differential equation
\begin{align}
 \frac{\rmd}{\rmd s} W_{(s,V)}^M(x) \equiv (\gLie_V W_{(s,V)})^M(x) \,,
\label{eq:genLie}
\end{align}
and obtain the finitely transformed generalized vector $W_{(s,V)}^M(x)$ for an arbitrary finite parameter $s$. 
We then rewrite the vector $W_{(s,V)}^M(x)$ in terms of the transformed coordinates, $x_s^M(x)$. 
It then becomes clear that the transformed coordinates $x_s^m(x)$ should be given by $x_s^m(x)=\Exp{s\,v}x^m$ in order to satisfy \eqref{eq:WsV-def}, while, since all fields are independent of the dual coordinates, the transformed dual coordinates $\tilde{x}^s_i(x)$ are arbitrary. 

Our key idea is to exploit the fact that the transformed dual coordinates $\tilde{x}^s_m(x)$ are arbitrary, equivalently, there is no canonical definition of them. 
Indeed, in DFT, as we shall recall in section \ref{sec:review-DFT}, several alternative definitions $\tilde{x}^s_m(x)=\Exp{s\,V}\tilde{x}_m$ or $\tilde{x}^s_m(x)=\tilde{x}_m$ were already considered.%
\footnote{If we realize the $B$-field gauge transformation as a diffeomorphism in the dual coordinates (as in Hohm and Zwiebach's approach \cite{Hohm:2012gk}), the three-form field strength, $H_3=\rmd B_2$, becomes globally an exact form due to the Papadopoulos problem \cite{Papadopoulos:2014mxa} (We would like to thank C.~Hull for conversation on this point). We comment on this issue in detail in section \ref{sec:dual-direction}.}
In section \ref{sec:dual-direction}, we put forward a new definition of $\tilde{x}^s_m(x)$ with which we can show that firstly the Papadopoulos problem does not arise and secondly the composition law is satisfied (see section \ref{sec:composition}). 
Our goal is to obtain the pullback $W_{(s,V)}^M(x)$ and express it in terms of the transformed coordinates $x_s^m(x)$ and a geometric quantity $b_{mn}(x)$ to be defined later. 

\section{Finite transformations in DFT}
\label{sec:finite-DFT}
In this section, we present details of our proposed definition of finite transformations in DFT. 
In section \ref{sec:review-DFT}, we begin by reviewing previous approaches for finite transformations in DFT. 
Since the doubled spacetime has an invariant tensor $\eta$, called the $\OO(d,d)$ metric, that is used to raise or lower the indices, once the finite transformation law for a generalized vector $W^M$ (or instead $W_M$) is obtained, the transformation law for arbitrary tensors can be readily obtained. 
The main task of this section is thus to obtain $W_{(s,V)}^M(x)$ by solving the differential equation \eqref{eq:genLie}. 
Our proposal for the finite transformation is defined in \ref{sec:Finite-geom}, which is physically intuitive in terms of the geometry of null subspaces. 
Our definition of the transformed dual coordinates and a discussion about the Papadopoulos problem is given in section \ref{sec:dual-direction}. 
The composition law of our finite transformations is shown in section \ref{sec:composition}. 
In section \ref{sec:non-geom}, we construct finite transformations in non-geometric backgrounds. 

\subsection{Review of previous studies}
\label{sec:review-DFT}

\subsubsection*{The Approach of Hohm and Zwiebach}
In the original approach by Hohm and Zwiebach \cite{Hohm:2012gk}, the finite transformation law (or pushforward) of a generalized vector field $W_M(x)$ under a (passive) finite coordinate transformation, $x^M\to x'^M=\Exp{-\sfV}x^M$, was proposed as 
\begin{align}
 W'_M(x') = \cF_M{}^N(x',x)\, W_N(x) \quad \text{where} \quad
\cF_M{}^N(x',x)\equiv \frac{1}{2}\,\biggl(\frac{\partial x'_M}{\partial x_P}\,\frac{\partial x^N}{\partial x'^P}+\frac{\partial x^P}{\partial x'^M}\,\frac{\partial x'_P}{\partial x_N}\biggr)\,. 
\end{align}
Imposing the strong constraint, they found that the transformation matrix $\cF_M{}^N$ obeys
\begin{align}
 \partial'_M=\cF_M{}^N\,\partial_N\,,\quad 
 \cF_M{}^K(x',x)\,\cF^N{}_K(x',x)=\delta_M^N\,,\quad 
 \cF_M{}^N(x,x')=\cF^N{}_M(x',x)\,.
\end{align}
They then showed that the transformed generalized vector $W'_M(x')$ can be expressed as the exponentiation of the generalized Lie derivative with a gauge parameter $V^M(x)$,
\begin{align}
 W'_M(x') \equiv \Exp{-\sfV}\cdot\, W'_M(x) 
  = \Exp{-V}\cdot\, W'_M(x) = \bigl(\Exp{-V}\Exp{\gLie_V}W\bigr)_M(x) \,. 
\label{eq:hohm-zwiebach-V-def}
\end{align}
Here, the gauge parameter, $V^M(x)$, is assumed to be related to the original parameter, $\sfV^M(x)$ (introduced for the coordinate change), via
\begin{align}
 \sfV^M(x) = V^M(x) + \sum_i \rho_i(x) \,\partial^M \chi_i(x) 
\end{align}
with certain functions $\rho_i(x)$ and $\chi_i(x)$ satisfying the strong constraint (which gives $\sfV^M\,\partial_M\,\circ = V^M\,\partial_M\,\circ$).\footnote{From the strong constraint, we have $\sum_i \rho_i(x) \,\partial^M \chi_i(x) \,\partial_M \phi(x)=0$ for arbitrary field $\phi(x)$, so from \eqref{eq:hohm-zwiebach-V-def}, we have $\sfV^M\,\partial_M \phi(x) = V^M\,\partial_M \phi(x)$. We thus obtain $\Exp{-\sfV} \cdot\ \phi(x) = \Exp{-V} \cdot\ \phi(x)$ and it shows the second equality in \eqref{eq:hohm-zwiebach-V-def}.}

In the coordinates with $\tilde{\partial}^m=0$, this relation has the form,
\begin{align}
 \sfV^M(x) =\begin{pmatrix} \sfv^m(x) \cr \tilde{\sfv}_m(x) \end{pmatrix}
 = \begin{pmatrix}
 v^m(x) \cr \tilde{v}_m(x) + \sum_i \rho_i(x) \,\partial_m \chi_i(x)\end{pmatrix} \, , 
\label{eq:hohm-zwiebach-cV}
\end{align}
so the difference between $V^M(x)$ and $\sfV^M(x)$ is only in the dual components. 
Thus, the standard components of the transformed coordinates is simply $x'^m(x)=\Exp{-\sfV}x^m=\Exp{-v}x^m$, as it should be. 
In \cite{Hohm:2012gk}, the explicit form of $\sum_i \rho_i(x) \,\partial^M \chi_i(x)$ satisfying the relation \eqref{eq:hohm-zwiebach-V-def} was found up to the quartic order in $V^M(x)$;
\begin{align}
 \sum_i \rho_i(x) \,\partial^M \chi_i(x) 
 = \frac{1}{12}\,\bigl(V\cdot V^N\bigr)(x)\,\partial^MV_N(x) + \cO(V^5) \,.
\end{align}
They also discussed in detail subtleties of their approach in the composition of finite transformations (see also \cite{Hohm:2013bwa} for further discussion). 
The higher order corrections for $\sum_i \rho_i(x) \,\partial^M \chi_i(x)$ within this approach was further studied in \cite{Naseer:2015tia}.

\subsubsection*{The Approach of Park}

In \cite{Park:2013mpa}, Park obtained a result similar to Hohm and Zwiebach from an active diffeomorphism point of view. 
As in the latter approach, the transformation matrix for a generalized vector field under a finite (active) diffeomorphism, $x^M \to x_s^M \equiv \Exp{sV}x^M$, was identified with the matrix $\cF_M{}^N$\,:
\begin{align}
 W_{sM}(x) \equiv \cF_M{}^N(x,x_s)\,W_N(x_s) \,.
\label{eq:Park-finite}
\end{align}
He showed that $\cF_M{}^N(x,x_s)$ satisfies 
\begin{align}
 &\frac{\rmd}{\rmd s}\cF_M{}^N 
 = \cV^K\,\partial_K \cF_M{}^N + \bigl(\partial_M \cV^K -\partial^K \cV_M\bigr)\,\cF_K{}^N \,,
\end{align}
where
\begin{align}
 &\cV^M \equiv V^M + \frac{1}{2}\, V_N\,\partial^M f_s^N \quad \text{and} \quad
 f_s^M \equiv \sum_{n=1} \frac{s^n}{n!}\,\bigl(V^N\,\partial_N\bigr)^{n-1} V^M \,. 
\label{eq:V-sfV-Park}
\end{align}
It follows that
\begin{align}
 \frac{\rmd}{\rmd s} W_{sM}(x) = \gLie_{\cV} W_{sM}(x) \, 
\end{align}
for an arbitrary $s$. 
Note that, since $\cV^M(x)$ explicitly depends on the parameter $s$, $W_{(s=1)M}(x)$ cannot be written as an exponential form $(\Exp{\gLie_\cV}W)_M(x)$. 
As such, it is not easy to compare the result with \eqref{eq:hohm-zwiebach-cV}. 

In \cite{Park:2013mpa}, coordinate transformations generated by generalized vectors of the form, $V^M=\phi^i\partial^M\varphi_i$, were called the coordinate gauge symmetry. 
As shown there, imposing the condition, $\tilde{\partial}^m=0$, this symmetry turns out equivalent to the gauge symmetry for the Kalb-Ramond two-form potential $B$. 
As the difference between the gauge parameter, $\cV^M(x)$, and the diffeomorphism parameter, $V^M(x)$, in \eqref{eq:V-sfV-Park} is just given by a coordinate gauge symmetry, it was claimed that the exponentiation of the generalized Lie derivative $(\Exp{s\,\gLie_V}W)_M(x)$ matches with \eqref{eq:Park-finite} up to the coordinate gauge symmetry. 
In \cite{Lee:2013hma}, it was further shown that if all points related by the coordinate gauge symmetry are identified as a single physical point, the strong constraint can be automatically derived. 
There also, a string worldsheet action which is invariant under the coordinate gauge symmetry was developed.

\subsubsection*{The approach by Berman, Cederwall, and Perry}

In this approach \cite{Berman:2014jba}, the finite transformation law under a passive transformation $x^M\to x'^M\equiv \Exp{-V}x^M$ was given by
\begin{align}
 W'_M(x') = (\Exp{-V}\Exp{\gLie_V}W)_M(x) \equiv \cG_M{}^N\,W_N(x) \,.
\label{eq:matrix-G-def}
\end{align}
They then showed that the transformation matrix, $(\cG_M{}^N)=\Exp{-V}\Exp{V+a-a^\mathrm{T}}$ ($a_M{}^N\equiv \partial_M V^N$), is related to $\cF_M{}^N$ via
\begin{align}
 \cF_M{}^N = (\cG \,\Exp{\Delta})_M{}^N 
\label{eq:F-G-Delta}
\end{align}
with
\begin{align}
\begin{split}
 &\Exp{\Delta} \equiv 
 \prod_{n=2}^\infty\prod_{k=0}^{n-1}
 \Bigl(1+\frac{1}{2}\frac{(-1)^n(n-2k-1)}{(n+1)(k+1)!(n-k)!}\,M_{n,k}\Bigr)\,,
\\
 &(M_{n,k})_M{}^N \equiv \partial_M \bigl(V^k\cdot V^L\bigr)\,\partial^N\bigl(V^{n-k-1}\cdot V_L\bigr) \,.
\end{split}
\end{align}
Under the constraint, $\tilde{\partial}^m=0$, the matrix $M_{n,k}$ has the only non-vanishing components, $(M_{n,k})_{mn}$, and $\Exp{\Delta}$ corresponds to a $B$-field transformation. 
They referred this $B$-field transformations as non-translating transformations, and claimed that the exponentiation of the generalized Lie derivative, $\cG_M{}^N$, matches with $\cF_M{}^N$ up to a non-translating transformation. They also confirmed consistency with Hohm and Zwiebach's result up to the cubic order in $V^M$. 

Moreover, they stressed that, if the matrix $\cF_M{}^N$ is regarded as a map, $\GL(2d)\to \OO(d,d)$, $M_M{}^N\equiv \partial_M x'^N\mapsto \cF_M{}^N(x',x)$, it is not a group homomorphism;
\begin{align}
 \cF(M)\,\cF(N)\neq \cF(MN) \quad \bigl(M,\,N\in\GL(2d)\bigr)\,. 
\end{align}
This implies the composition law is not satisfied. 
They further discussed the relation between the failure of the composition law and the gerbe structure associated with the $B$-field transformations. 

\subsubsection*{The approach of Hull}

A different approach, using an untwisted vector, was developed in \cite{Hull:2014mxa}. 
In this approach, the strong constraint $\tilde{\partial}^m=0$ is solved at the outset, and all 
fields depend only on the coordinates $x^m$ of a $d$-dimensional null subspace $\cN$. 
Further, one considers the coordinate change only on the null plane $\cN$ (which is one leaf of a constant-$\tilde{x}_m$ foliation), while the dual coordinates are always kept invariant, $\tilde{x}'_m = \tilde{x}_m$. 
Now, for a given generalized vector $W^M(x)$, one associates the untwisted vector, $\widehat{W}^M(x)$, by
\begin{align}
 \widehat{W}^M(x) \equiv \begin{pmatrix} w^m(x) \cr \widehat{w}_m(x) \end{pmatrix}
  \equiv \begin{pmatrix} w^m(x) \cr \widetilde{w}_m(x) - B_{mn}(x)\,w^n(x) \end{pmatrix}
  = \begin{pmatrix} \delta^m_n  & 0 \cr -B_{mn}(x) & \delta_m^n \end{pmatrix}
    \begin{pmatrix} w^n(x) \cr \widetilde{w}_n(x) \end{pmatrix} \,.
\label{eq:untwisted-vector}
\end{align}
Here, $B_{mn}(x)$, called a gerbe connection, is an anti-symmetric two-form which transforms as
\begin{align}
 \delta_V B_{mn}(x) = \Lie_v B_{mn}(x) + \partial_m\widetilde{v}_n(x)-\partial_n\widetilde{v}_m(x) \,,
\label{eq:transformation-B-inf}
\end{align}
under the gauge transformation. 
In \cite{Hull:2014mxa}, it was chosen to be the Kalb-Ramond $B$-field. 

An important property of the untwisted vector $\widehat{W}^M$ is that the transformation under the gauge symmetry is simply given by the conventional Lie derivative:
\begin{align}
 \delta_V\widehat{W}^M(x) 
 = \begin{pmatrix} \Lie_v w^m(x) \cr \Lie_v \widehat{w}_m(x) \end{pmatrix} \,.
\label{eq:untwisted-Lie}
\end{align}
As such, the untwisted vector $\widehat{W}^M(x)$ is independent of transformations generated by $\tilde{v}_m(x)$. 
Therefore, the transformation law under a finite (passive) coordinate transformation, $x^m\to x'^m$, is given by
\begin{align}
 \widehat{W}^M(x') 
 = \begin{pmatrix} \frac{\partial x'^m}{\partial x^n} & 0 \cr 0 & \frac{\partial x^n}{\partial x'^m} \end{pmatrix}
   \begin{pmatrix} w^n(x) \cr \widetilde{w}_n(x) \end{pmatrix} \,.
\label{eq:finite-untwisted}
\end{align}
Under the constraint $\tilde{\partial}^m=0$, the gauge symmetry of DFT is a semi-direct product of diffeomorphisms and $B$-field transformations, so a finite transformation law for the $B$-field should be expressible as
\begin{align}
 B'_{mn}(x') = \frac{\partial x^p}{\partial x'^m}\,\frac{\partial x^q}{\partial x'^n}\, \bigl(B_{pq} + \partial_p \fv_q -\partial_q \fv_p\bigr)(x) \,,
\label{eq:transformation-b-finite}
\end{align}
where $\fv_m(x)$ is a certain gauge parameter associated with the finite $B$-field gauge transformation. 
By combining \eqref{eq:finite-untwisted} and \eqref{eq:transformation-b-finite}, the transformation law for the generalized vector $W^M(x)$ is obtained as
\begin{align}
 W'^M(x') 
 &= \begin{pmatrix} \delta^m_n & 0 \cr B'_{mn}(x') & \delta_m^n \end{pmatrix}
    \begin{pmatrix} w'^n(x') \cr \widehat{w}'_n(x') \end{pmatrix} 
  = \cR^M{}_N\,W^N(x) \,,
  \end{align}
  where
  \begin{align}
 \cR^M{}_N&\equiv \begin{pmatrix} \delta^m_k & 0 \cr B'_{mk}(x') & \delta_m^k \end{pmatrix}
\begin{pmatrix} \frac{\partial x'^k}{\partial x^l} & 0 \cr 0 & \frac{\partial x^l}{\partial x'^k} \end{pmatrix}
    \begin{pmatrix} \delta^l_n & 0 \cr -B_{ln}(x) & \delta_l^n \end{pmatrix}
\nn\\
 &= \begin{pmatrix} \frac{\partial x'^m}{\partial x^k} & 0 \cr 0 & \frac{\partial x^k}{\partial x'^m} \end{pmatrix}
    \begin{pmatrix} \delta^k_n & 0 \cr 2\,\partial_{[k}\fv_{n]}(x) & \delta_k^n \end{pmatrix} \,.
\label{eq:Hull-finite}
\end{align}
This transformation matrix is different form the transformation matrix $\cF^M{}_N$ (restricted to $\tilde{\partial}^m=0$) for the coordinate transformation, $x'^m=\Exp{-v}x^m$ and $\widetilde{x}'_m=\widetilde{x}_m-\zeta_m(x)$ (see (2.42) in \cite{Hohm:2013bwa}):
\begin{align}
 \cF^M{}_N = \begin{pmatrix} \delta^m_k & 0 \cr \partial'_{[m}\zeta_{k]} (x) & \delta_m^k \end{pmatrix}
 \begin{pmatrix} \frac{\partial x'^k}{\partial x^l} & 0 \cr 0 & \frac{\partial x^l}{\partial x'^k} \end{pmatrix}
    \begin{pmatrix} \delta^l_n & 0 \cr \partial_{[l}\zeta_{n]}(x) & \delta_l^n \end{pmatrix} \, . 
\label{eq:H-Z-F}
\end{align}
On the other hand, the two transformation matrices are equivalent up to a $B$-field gauge transformation. 

Given three coordinate transformations with transition functions, $\cR_1$ $(U_\alpha\to U_\beta)$, $\cR_2$ $(U_\beta\to U_\gamma)$ and $\cR_3$ $(U_\gamma\to U_\alpha)$ on a triple overlap of patches, $U_\alpha\cap U_\beta \cap U_\gamma$, it can be shown that the cocycle condition is automatically satisfied, $\cR_1\, \cR_2 \, \cR_3=1$ \cite{Hull:2014mxa}. 
If one instead use Hohm and Zwiebach's $\cF$ as the transition function, one obtains $\cF_1\cF_2\cF_3\neq 1$ and finds a gerbe-like structure \cite{Berman:2014jba}, implying violation of the composition law. 
Thus, there is no reason to consider the matrix $\cF_M{}^N$ as an exact expression for the finite transformation matrix; $\cF_M{}^N$ is equal to the exponential of the generalized Lie derivative (or $G_M{}^N$ in \eqref{eq:matrix-G-def}) only up to a $B$-field gauge transformation and also does not satisfy the composition law.

It is worth to note that if we identify all spacetime points related by the coordinate gauge symmetry as in \cite{Park:2013mpa}, the transformation matrices $\cF_M{}^N$ and $\cG_M{}^N$ are equivalent each other. 
In this case, one can show that both of them satisfies the composition law since all $B$-field gauge transformations can be canceled by using the coordinate gauge symmetry. 
Because of this, one might consider the approach of \cite{Park:2013mpa} is consistent. 
However, assumption of the coordinate gauge symmetry is restrictive since, after modding out the $B$-field transformations, the remaining gauge symmetry is only the $d$-dimensional diffeomorphisms and we cannot consider a non-trivial patching of doubled spacetime \cite{Hull:2014mxa}. 
In this sense, the approach of \cite{Hull:2014mxa} is more satisfactory, as the finite transformations obey the composition law without modding out $B$-field transformations.

It was shown in \cite{Hull:2014mxa} that, under the condition $\tilde{\partial}^m=0$, the finite transformation matrix can always be expressed in the form \eqref{eq:Hull-finite}. 
On the other hand, it is also the case that finite transformation can be expressed as the exponential of generalized Lie derivative, $W'^M(x')= (\Exp{-V}\Exp{\gLie_V}W)_M(x)$, which was the original problem posed in \cite{Hohm:2012gk,Park:2013mpa,Berman:2014jba,Naseer:2015tia}. 
In the decomposition of the gauge parameter as $V^M=(v^m,\,\widetilde{v}_m)$, the parameter $v^m$ would be related to $x'^m$ in \eqref{eq:Hull-finite} by $x'^m=\Exp{-v}x^m$. On the other hand, the relation between $\fv_m(x)$, that was introduced in \eqref{eq:transformation-b-finite}, and $V^M(x)$ was not identified in \cite{Hull:2014mxa}. 
Since a finite transformation of the $B$-field is obtained by integrating the infinitesimal transformation \eqref{eq:transformation-B-inf}, $\fv_m(x)$ should be expressible as a complicated function of $V^M(x)$. 
In this paper, we give an affirmative answer to this question. 
This is an important step since, as we demonstrate below in section \ref{sec:composition}, a proof of the composition law requires the explicit form of $\fv_m(x)$, which fell short in the analysis of \eqref{eq:Hull-finite}.%
\footnote{We can of course show that two consecutive  finite transformations, generated by $x^m\to x_1^m$ and $\fv_m^1$, and $x_1^m\to x_2^m$ and $\fv_m^2$, can be composed into a single transformation, $x^m\to x_2^m$ and $\fv_m^{12}$ with $\fv_m^{12}$ a certain gauge parameter. However, in the absence of the explicit form of $\fv_m$, we can only claim the following trivial statement: a composition of two finite diffeomorphisms and $B$-field gauge transformations can be expressed again as a single finite diffeomorphisms and a $B$-field gauge transformation.} 

\subsection{Our approach: finite transformations in geometric backgrounds}
\label{sec:Finite-geom}
We now construct a new finite transformation law, in geometric backgrounds in this subsection and in non-geometric backgrounds in subsection 3.4. 
We use the untwisted vector as in the approach of \cite{Hull:2014mxa}. 
However, the key difference of our approach is that we express the finite transformation for the $B$-field \eqref{eq:transformation-b-finite} in terms of the transformed coordinates, $x_s^m(x)=\Exp{s\,v}x^m$, and a `geometric' quantity, $b_{mn}(x)$, instead of the gauge parameter, $\fv_m(x)$. 
We begin with defining the quantity $b_{mn}(x)$ and obtain its finite transformation law. 
Using this, we then show that we can readily obtain the finite transformation law for the field $B_{mn}(x)$ in a form similar to \eqref{eq:transformation-b-finite}. 
Combining the finite transformation laws for the untwisted vector $\widehat{W}^M(x)$ and the field $B_{mn}(x)$, we finally construct the finite transformations for the generalized vector, $W_{(s,V)}^M(x)$. 

In our proposed approach, the gauge parameter $V^M(x)$ is $s$-independent. 
Therefore, the transformed vector and the conventional coordinates can be expressed by exponential maps as $W_{(s,V)}^M(x)=(\Exp{s\gLie_V}W)^M(x)$ and $x_s^m(x)=\Exp{s\,v}x^m$. 
On the other hand, the dual coordinates in our approach do not obey an exponential map like $\tilde{x}^s_m(x)=\Exp{s\,V}\tilde{x}_m$. 
Recall that, as we discussed in section \ref{sec:finite-extended}, the generalized coordinates do not need to transform as $\delta_V x^M=V^M$ under a gauge transformation, $\delta_V W^M=\gLie_V W^M$. 

\subsubsection*{Definition of $b_{mn}$}

In order to formulate the definition of $b_{mn}(x)$, we will first need to make some geometric setups. 
We introduce local patch of generalized coordinates, $x^M=(x^m,\,\widetilde{x}_m)$, and demand the constraint $\tilde{\partial}^m=0$ patchwise on all fields. 
In this coordinate patch, we choose a foliation of the doubled spacetime $\cM$ by $d$-dimensional null surfaces and take a leaf $\cN$ to correspond to a physical subspace (on which physical fields are defined). 
We also denote a set of $d$ tangent vector fields on the null surface by $\{e_a(x)\}$ ($a=1,\dotsc,d$). 
The frame $\{e_a(x)\}$ should be linearly independent from the dual frame $\{\widetilde{e}^a\equiv \widetilde{\partial}^a\}$, so we parameterize them as
\begin{align}
\boxed{~
 e^M{}_a(x)= \begin{pmatrix} \delta^m_a \cr b_{ma}(x) \end{pmatrix}\qquad \text{viz.}\qquad
 e_a(x) = \partial_a - b_{am}(x)\,\widetilde{\partial}^m \,,~
 }
\end{align}
for some field $b_{am}(x)$. 
Here, $b_{am}(x)$ has to be anti-symmetric since, from the strong constraint, $\cN$ should be a null surface with respective to $\eta_{MN}$:
\begin{align}
 0=\eta_{MN}\,e^M{}_a\,e^N{}_b = b_{ab} + b_{ba} \,.
\label{eq:DFT-null}
\end{align}
We can reorganize the frame fields, $\{e_a\}$ and $\{\widetilde{e}^a\}$, into a $2d$-dimensional generalized frame field:
\begin{align}
 E^M{}_A(x) \equiv (e^M{}_a(x),\,\widetilde{e}^{Ma}(x))=\begin{pmatrix} \delta^m_a & 0 \cr b_{ma}(x) & \delta_m^a \end{pmatrix} \quad \text{viz.} \quad 
 E_A(x)=(\partial_a - b_{am}(x)\,\widetilde{\partial}^m,\,\widetilde{\partial}^a)\,.
\end{align}
Under a gauge transformation (generated by the generalized Lie derivative), this frame field may change its form. In order to keep the form as $e^m{}_a=\delta^m_a$ and $\widetilde{e}_m{}^a=\delta_m^{a}$, we also make a $\GL(d)$ transformation.%
\footnote{See discussion around (5.53) in \cite{Hohm:2010pp} for similar reasonings.} 
The combined transformation reads
\begin{align}
 \delta_V E^M{}_A
 = \bigl[\, V^N\,\partial_N E^M{}_A - \bigl(\partial_N V^M-\partial^M V_N\bigr)\,E^N{}_A \, \bigr] 
  +E^M{}_B\, \begin{pmatrix} \partial_a v^b & 0 \cr 0 & -\partial_b v^a \end{pmatrix}\,.
\label{eq:delta-E}
\end{align}
With $V^M = (v^m, \widetilde{v}_m)$, this gives the transformation law for $b_{mn}(x)$,
\begin{align}
 \delta_V b_{mn}(x) = \Lie_v b_{mn}(x)+\partial_m \tilde{v}_n(x) - \partial_n \tilde{v}_m(x) \,.
\label{eq:transformation-b-inf}
\end{align}
Note that, with the gauge conditions $e^m{}_a=\delta^m_a$ and $\tilde{e}_m{}^a=\delta_m^{a}$, there is no difference between the index $a$ with the index $m$. 

If we expand a generalized vector as $W^M = E^M{}_A\,\mathsf{W}^A$ with $\mathsf{W}^A\equiv (w^a,\,\widehat{\mathsf{w}}_a)$, from \eqref{eq:delta-E}, its generalized Lie derivative can be written as
\begin{align}
 \gLie_V W^M(x) = E^M{}_A(x)\,\delta_V \mathsf{W}^A(x) + \delta_V E^M{}_A(x)\,\mathsf{W}^A(x)\,,
\end{align}
where
\begin{align}
 \delta_V \mathsf{W}^A(x) =\begin{pmatrix} (\Lie_v w)^a(x) \cr (\Lie_v \widehat{\mathsf{w}})_a(x) \end{pmatrix} \qquad \text{and} \qquad 
 \delta_V E^M{}_A(x) = \begin{pmatrix} 0 & 0 \cr \delta_V b_{ma}(x) & 0 \end{pmatrix}\,.
\label{eq:genLie-decomp}
\end{align}
We can view this as a decomposition of the generalized Lie derivative into the conventional Lie derivative on the $d$-dimensional space $\cN$ and the deformation of $\cN$.%
\footnote{Note that this decomposition is similar to (4.4) in \cite{Asakawa:2012px}.} 
This interpretation is one of main thesis of our approach. 

Further, we require that, under an arbitrary diffeomorphism of the null surface $\cN$, the tangent vector fields on the surface $\cN$ remains to be tangent. 
Namely, we require that a variation of an arbitrary tangent vector $W^M=e^M{}_a\,w^a \in T\cN$ under a gauge transformation generated by $V^M=e^M{}_a\,v^a\in T\cN$ is again spanned by $e^M{}_a$. 
Then, since the generalized Lie derivative in this case becomes
\begin{align}
 \bigl(\gLie_V W\bigr)^M = e^M{}_a \,\bigl(\Lie_v w\bigr)^a 
 + \begin{pmatrix} 0 \cr 3\,\partial_{[m}b_{nl]} \end{pmatrix}\,w^n\,v^l \,,
\end{align}
and the second term cannot be expanded by $e^M{}_a$, the following condition is required:
\begin{align}
 \partial_{[m} b_{nl]}(x) =0 \,.
\label{eq:involutive}
\end{align}

\subsubsection*{Finite transformation law}

We now construct the finite transformation law for $b_{mn}(x)$ from the infinitesimal transformation \eqref{eq:transformation-b-inf}. 
More specifically, denoting the finitely transformed $b_{mn}(x)$ by $b^{(s,V)}_{mn}(x)$, we do so by solving the differential equation,
\begin{align}
 \frac{\rmd}{\rmd s}b^{(s,V)}_{mn}(x) 
 = \Lie_v b^{(s,V)}_{mn}(x) + \partial_m\widetilde{v}_n(x)-\partial_n\widetilde{v}_m(x) \,.
\label{eq:b_s-diff}
\end{align}
By expanding $b^{(s,V)}_{mn}(x)$ in power series
\begin{align}
 b^{(s,V)}_{mn}(x) = b_{mn}(x) + \sum_{k=1}^\infty \frac{s^k}{k!}b^{[k]}_{mn}(x) \,,
\end{align}
we easily find the formal solution,
\begin{align}
 b^{(s,V)}_{mn}(x) = b_{mn}(x) + \sum_{k=1}^\infty \frac{s^k}{k!}\,\bigl[\Lie_v^{k-1}\bigl(\Lie_v b+\rmd \widetilde{v}\bigr)\bigr]_{mn} \,,
\label{eq:formal-b}
\end{align}
where we treated $\widetilde{v}_m(x)$ as a one-form. 
One way of rewriting this solution is as follows:
\begin{align}
 b^{(s,V)}_{mn}(x) &= \,  \,   \Exp{s\Lie_v} b_{mn}(x) \, \, 
 + \, \, \biggl[\rmd \sum_{k=1}^\infty \frac{s^k}{k!}\,\Lie_v^{k-1} \widetilde{v}\biggr]_{mn} 
\nn\\
 &=\frac{\partial x_s^k}{\partial x^m}\,\frac{\partial x_s^l}{\partial x^n}\,b_{kl}(x_s) 
 + \biggl[\rmd \sum_{k=1}^\infty \frac{s^k}{k!}\,\Lie_v^{k-1} \widetilde{v}\biggr]_{mn} \,,
\end{align}
where $x_s^m\equiv \Exp{s\,v}x^m$ is the transformed conventional coordinates. 

Since $B^{(s,V)}_{mn}(x)$ also satisfies the same differential equation \eqref{eq:b_s-diff}, the above form of finite transformation law must also be valid for the $B$-field:
\begin{align}
 B^{(s,V)}_{mn}(x) = \frac{\partial x_s^k}{\partial x^m}\,\frac{\partial x_s^l}{\partial x^n}\,B_{kl}(x_s) 
 + \biggl[\rmd \sum_{k=1}^\infty \frac{s^k}{k!}\,\Lie_v^{k-1} \widetilde{v}\biggr]_{mn} \,. 
\label{eq:B-finite-tilde}
\end{align}
However,  we consider the one-form field, 
\begin{align}
\label{sfv}
 \widehat{\sfv}_m(x) \equiv (\widetilde{v}+\iota_v b)_m(x) = \widetilde{v}_m(x) - b_{mn}(x)\,v^n(x) \,,
\end{align}
more appropriate than $\widetilde{v}_m(x)$ and rewrite the transformation law in terms of this field. 
This is because $\widehat{\sfv}_m(x)$ is invariant under the $B$-field transformations and transforms as a conventional one-form, similar to the dual components of the untwisted vector (although a different gerbe connection, $b_{mn}(x)$, is used here). 
Using \eqref{eq:involutive}, $(\rmd b)_{mnl}=0$, one can show that
\begin{align}
 \delta_V b_{mn} = \bigl(\Lie_v b+\rmd \widetilde{v}\bigr)_{mn}
 = \bigl(\rmd\iota_v b+\rmd \widetilde{v}\bigr)_{mn}
 = (\rmd \widehat{\sfv})_{mn}\,,
\end{align}
and that the finite transformation \eqref{eq:formal-b} can also be written as
\begin{align}
\label{eq:b_s-sol}
 b^{(s,V)}_{mn}(x) &= b_{mn}(x) + \sum_{k=1}^\infty \frac{s^k}{k!}\,\bigl(\Lie_v^{k-1}\rmd \widehat{\sfv}\bigr)_{mn} 
 = b_{mn}(x) + 2\,\partial_{[m} \zeta^{(s,V)}_{n]}(x) \,,
\end{align}
where
\begin{align}
\begin{split}
 \zeta^{(s,V)}_m(x)&\equiv \sum_{k=1}^\infty \frac{s^k}{k!}\,\bigl(\Lie_v^{k-1} \widehat{\sfv}\bigr)_m(x)
 = \int_0^s\rmd s'\, \widehat{\sfv}^{(s',V)}_m(x) \,,
\\
 \hat{\sfv}^{(s,V)}_m(x) 
 &\equiv \Exp{s\Lie_v}\widehat{\sfv}_m(x) = \frac{\partial x_s^n}{\partial x^m}\, \widehat{\sfv}_n(x_s) \,.
\end{split}
\end{align}
This shows that $b_{mn}(x)$ changes by a locally exact form under an arbitrary finite transformation that is connected to the identity. 
In particular, under a finite transformation along a trivial Killing vector (which was referred as a trivial parameter in \cite{Hohm:2012gk}) with the form, $V^M=\partial^Mf(x)=(0,\,\partial_m f)$, where $f(x)$ is an arbitrary function satisfying $\widetilde{\partial}^mf(x)=0$, $\zeta_m^{(s,V)}(x)$ becomes a (locally) exact one-form and $b_{mn}(x)$ is not changed. 
In our approach, we will identify two gauge parameters which differ by a trivial Killing vector. 
Then, $\zeta^{(s,V)}_m$, $\widehat{\sfv}_m$, and $\widehat{\sfv}^{(s,V)}_m$ are all defined up to a (locally) exact one-form.%
\footnote{When we count the number of the off-shell degrees of freedom of the $B$-field, we subtract from $\frac{d(d-1)}{2}$ a number of independent gauge parameters associated with the gauge symmetry, $B_{mn}\to B_{mn}+2\,\partial_{[m} \widetilde{v}_{n]}$. Although, the gauge parameter $\widetilde{v}_m$ has $d$ components, since it is defined only up to a equivalence relation, $\widetilde{v}_m\simeq \widetilde{v}_m+\partial_m f$, the number of the degrees of freedom of the $B$-field becomes $\frac{1}{2} d(d-1)-(d-1)=\frac{1}{2} (d-1)(d-2)$. The equivalence relation here corresponds to regarding the trivial Killing vector as a zero vector.} 

In \eqref{sfv}, we defined the untwisted vector field $\widehat{\sfv}_m(x)$ using the gerbe connection $b_{mn}(x)$ before the finite transformation. 
Since it behaves as a conventional one-form, after a finite transformation, it becomes $\widehat{\sfv}^{(s,V)}_m(x)=\Exp{s\Lie_v}\widehat{\sfv}_m(x)$. 
On the other hand, we can also define the untwisted vector field $\widehat{\sfv}_m(x)$ after the finite transformation. 
Namely, after the finite transformation, the generalized vector $V^M(x)$ becomes $V_{(s,V)}^M(x)\equiv \bigl(v_{(s,V)}^m,\,\widetilde{v}^{(s,V)}_m\bigr)\equiv \Exp{s\gLie_V}V^M(x)=\bigl(v^m(x),\,\widetilde{v}_m(x) + \sum_{k=1}^\infty\frac{s^k}{k!}\,\bigl(\rmd v^{k-1} \iota_v\widetilde{v}\bigr)_m(x)\bigr)$. So, we can define an untwisted vector $\lambda_m^{(s,V)}$ after the finite transformation as
\begin{align}
 \lambda^{(s,V)}_m(x)&\equiv \bigl[ \widetilde{v}^{(s,V)}+\iota_{v_{(s,V)}} b^{(s,V)}\bigr]_m(x) 
\nn\\
 &= \widetilde{v}_m(x) + \sum_{k=1}^\infty\frac{s^k}{k!}\,\bigl(\rmd v^{k-1}\cdot\iota_v\widetilde{v}\bigr)_m(x) - b^{(s,V)}_{mn}(x)\,v^n(x) \,.
\end{align}
Using the solution \eqref{eq:b_s-sol}, one finds
\begin{align}
 \lambda^{(s,V)}_m(x) = \widehat{\sfv}_m(x) + \sum_{k=1}^\infty\frac{s^k}{k!}\,\bigl(\Lie_v^k\widehat{\sfv}\bigr)_m(x)
  = \Exp{s\Lie_v}\widehat{\sfv}_m(x) = \widehat{\sfv}^{(s,V)}_m(x) \,.
\label{eq:lambda-v}
\end{align}
Therefore, the untwisting procedure, which maps a generalized vector $V^M= (v^m,\,\tilde{v}_m)$ to an untwisted vector $\mathsf{V}^A= (v^a,\,\widehat{\mathsf{v}}_a)$, and the finite transformation commute, and the untwisting by the gerbe connection $b_{mn}$ is a well-defined procedure. 

Regarding $\zeta^{(s,V)}_m(x)$ as a one-form, we can express the finite transformation \eqref{eq:b_s-sol} as
\begin{align}
 b^{(s,V)}_{mn}(x) = (b+\rmd \zeta^{(s,V)})_{mn}(x) \,. 
\label{eq:finite-bmn}
\end{align}
Using this finite transformation law for $b_{mn}(x)$, we can easily obtain the finite transformation for the $B$-field. 
Recalling the infinitesimal transformation law,
\begin{align}
 \delta_V B_{mn}(x) = \Lie_v B_{mn}(x)+\partial_m \widetilde{v}_n(x) - \partial_n \widetilde{v}_m(x) \,,
\end{align}
one can show the combination, $\bB_{mn}(x)\equiv B_{mn}(x)-b_{mn}(x)$, transforms as
\begin{align}
 \delta_V \bB_{mn}(x) = \Lie_v \bB_{mn}(x) \,.
\end{align}
Therefore, the finite transformation is simply given by
\begin{align}
 \bB^{(s,V)}_{mn}(x) = \frac{\partial x_s^k}{\partial x^m}\,\frac{\partial x_s^l}{\partial x^n}\,\bB_{kl}(x_s) \,.
\end{align}
Then, using \eqref{eq:finite-bmn}, we obtain the finite transformation for the $B$-field:%
\footnote{This is consistent with \eqref{eq:B-finite-tilde}, as can be shown using $(\rmd b)_{kmn}=0$\,:
\begin{align*}
 B^{(s,V)}_{mn} 
 &= \Exp{s\Lie_v} (B_{mn} - b_{mn}) + b_{mn} + (\rmd \zeta^{(s,V)})_{mn} 
\\
 &= \Exp{s\Lie_v} B_{mn} -\Bigl(\rmd \sum_{k=1}\frac{s^k}{k!} \Lie_v^{k-1} \iota_v b\Bigr)_{mn} + (\rmd \zeta^{(s,V)})_{mn} 
  = \Exp{s\Lie_v} B_{mn} +\Bigl(\rmd \sum_{k=1}\frac{s^k}{k!} \Lie_v^{k-1} \widetilde{v}\Bigr)_{mn} \,.
\end{align*}
}
\begin{align}
\boxed{~
 B^{(s,V)}_{mn}(x) = \frac{\partial x_s^k}{\partial x^m}\,\frac{\partial x_s^l}{\partial x^n}\,\bigl(B_{kl} -b_{kl}\bigr)(x_s) 
+ b_{mn}(x) + 2\,\partial_{[m}\zeta^{(s,V)}_{n]}(x) \,.~
}
\label{eq:B-finite}
\end{align}
We emphasize that the finite transformation depends not only gauge parameters but also local specification of the null surface $\cN$ by $b_{mn}(x)$, which can be chosen arbitrarily.

Combining the above results, we obtain the finite transformation for a generalized vector field:
\begin{align}
 W_{(s,V)}^M(x)
 &= \begin{pmatrix} \delta^m_k & 0 \cr B^{(s,V)}_{mk}(x) & \delta_m^k \end{pmatrix}
\begin{pmatrix} \frac{\partial x^k}{\partial x_s^l} & 0 \cr 0 & \frac{\partial x_s^l}{\partial x^k} \end{pmatrix}
 \begin{pmatrix} \delta^l_n & 0 \cr -B_{ln}(x_s) & \delta_l^n \end{pmatrix}
 \begin{pmatrix} w^n(x_s) \cr \widetilde{w}_n(x_s) \end{pmatrix}
\nn\\
 &= \begin{pmatrix} \delta^m_k & 0 \cr b^{(s,V)}_{mk}(x) & \delta_m^k \end{pmatrix}
\begin{pmatrix} \frac{\partial x^k}{\partial x_s^l} & 0 \cr 0 & \frac{\partial x_s^l}{\partial x^k} \end{pmatrix}
 \begin{pmatrix} \delta^l_n & 0 \cr -b_{ln}(x_s) & \delta_l^n \end{pmatrix}
 \begin{pmatrix} w^n(x_s) \cr \widetilde{w}_n(x_s) \end{pmatrix} \,.
\label{eq:finite-transf}
\end{align}
Since each matrix is an $\OO(d,d)$ element, the whole transformation matrix is also an $\OO(d,d)$ element. 
Defining our transformation matrix,
\begin{align}
\boxed{~
 \ours^M{}_N \equiv \begin{pmatrix} \delta^m_k & 0 \cr b^{(s,V)}_{mk}(x) & \delta_m^k \end{pmatrix}
\begin{pmatrix} \frac{\partial x^k}{\partial x_s^l} & 0 \cr 0 & \frac{\partial x_s^l}{\partial x^k} \end{pmatrix}\begin{pmatrix} \delta^l_n & 0 \cr -b_{ln}(x_s) & \delta_l^n \end{pmatrix} \,,~
}
\label{eq:cR-DFT}
\end{align}
the finite transformation law for an arbitrary $(k,l)$-tensor field can be written as%
\footnote{For a tensor density of the weight $w$, there is an additional factor $[\det (\partial x_s^m/\partial x^n)]^w$ on the right-hand side.}
\begin{align}
 (T_{(s,V)})^{M_1\cdots M_k}_{N_1\cdots N_l}(x)
 = \ours^{M_1}{}_{K_1}\cdots \ours^{M_k}{}_{K_k}\,\ours_{N_1}{}^{L_1}\cdots \ours_{N_l}{}^{L_l}\,T^{K_1\cdots K_k}_{L_1\cdots L_l}(x_s) \,.
\label{eq:our-formula}
\end{align}
Once again, the transformation depends not only the diffeomorphisms but also local specification of the null surface $\cN$ by $b_{mn}(x)$.

As emphasized above, the two-form field, $b_{mn}(x)$, specifies arbitrary embedding of the initial null surface $\cN$ inside the doubled spacetime. This then introduces an arbitrariness in the expression of the transformation matrix $\ours^M{}_N$. In particular, if we choose the initial null surface $\cN$ as a constant-$\tilde{x}_m$ surface, we have $b_{mn}=0$ and the transformation matrix $\ours^M{}_N$ becomes
\begin{align}
 \ours^M{}_N &= \begin{pmatrix} \delta^m_k & 0 \cr 2\,\partial_{[m}\zeta^{(s,V)}_{k]}(x) & \delta_m^k \end{pmatrix}
\begin{pmatrix} \frac{\partial x^k}{\partial x_s^n} & 0 \cr 0 & \frac{\partial x_s^n}{\partial x^k} \end{pmatrix}
\nn\\
 &= \begin{pmatrix} \frac{\partial x^m}{\partial x_s^k} & 0 \cr 0 & \frac{\partial x_s^k}{\partial x^m} \end{pmatrix}
 \begin{pmatrix} \delta^k_n & 0 \cr 2\,\partial_{[k}\zeta^{(s,V)}_{n]}(x_s) & \delta_k^n \end{pmatrix}\,.
\end{align}
If one compares this with the finite transformation matrix ${\cR^M}_N$ of \eqref{eq:Hull-finite} in Hull's approach, one observes that the two are related each other by the relation $\rmd \fv=\rmd \zeta^{(s=1,V)}$.%
\footnote{This comparison may not be so meaningful since in our approach $b_{mn}$ changes under the generalized diffeomorphisms while in Hull's approach $b_{mn}$ is always set to zero. The two approaches are just different.} 

We now offer a physical interpretation of our transformation matrix \eqref{eq:cR-DFT}. 
The rightmost factor in \eqref{eq:cR-DFT} represents the untwisting procedure. 
This factor is a part of the non-translating transformation and does not change the physical coordinates, $x^m$. However, it transforms the initial foliation into the foliation by null planes, constant-$\widetilde{x}_m$ surfaces. 
The middle factor represents the conventional diffeomorphism on the $d$-dimensional null plane $\cN$ with $b_{mn}=0$. 
The leftmost factor represents the twisting procedure with $b^{(s,V)}_{mn}$, which makes the foliation by null planes into a foliation by null surfaces characterized by $b^{(s,V)}_{mn}$ (i.e.~the inverse of the untwisting procedure). 
Note that the transformation \eqref{eq:finite-transf} can be understood as the finite version of \eqref{eq:genLie-decomp}; the diffeomorphism is a finite version of $\Lie_v$ and the (un)twisting matrices are finite versions of $\delta_V E^M{}_A$. 
Further, if we define an ``untwisted'' generalized vector, $\mathsf{V}^A\equiv (v^a,\,\hat{\mathsf{v}}_a)$ by $V^M=E^M{}_A\,\mathsf{V}^A$ (much like the decomposition of $W^M$ in \eqref{eq:genLie-decomp}), the identity, $\delta_V b_{mn} = (\rmd \hat{\sfv})_{mn}$, implies that the upper component, $v^a(x)$, corresponds to a gauge parameter for a diffeomorphism along a null surface $\cN$ which does not deform the foliation. 
On the other hand, the lower component, a one-form $\widehat{\mathsf{v}}_a(x)$, does not generate a diffeomorphism along the null surface but change the foliation. 

\subsection{Diffeomorphisms along the dual direction and Papadopoulos problem}
\label{sec:dual-direction}

So far, we did not consider the diffeomorphisms in dual directions explicitly. In general, the generalized diffeomorphisms along the dual direction induces a deformation of the null subspace $\cN$ and hence a change to the field $b_{mn}$. 
We here comment on subtle issues related to the diffeomorphisms along dual directions. 

As we reviewed in section \ref{sec:review-DFT}, in the original approach by Hohm and Zwiebach \cite{Hohm:2012gk}, the finite transformation with the constraint, $\widetilde{\partial}^m=0$, was given by
\begin{align}
W'_M(x') = \cF_M{}^N(x',x)\, W_N(x) \quad \text{with} \quad x'^m =x'^m(x)\,,\quad \tilde{x}'_m = \widetilde{x}_m - \zeta_m(x) \,.
\end{align}
Adopting this finite transformation law as a patching condition between two local patches, $\cU_\alpha$ and $\cU_\beta$, we obtain
\begin{align}
\begin{split}
 &W_{(\alpha)M}(x_{(\alpha)}) = \cF_M{}^N(x_{(\alpha)},x_{(\beta)})\, W_{(\beta)N}(x_{(\beta)}) 
\\
\text{with} \qquad &x_{(\alpha)}^m = x^m_{(\alpha\beta)}(x_{(\beta)})\,,\quad \widetilde{x}_{(\alpha)m} = \widetilde{x}_{(\beta)m} - \zeta_{(\alpha\beta)m} \,,
\end{split}
\label{eq:patch-generalized-coords}
\end{align}
where $x^m_{(\alpha)}$ and $x^m_{(\beta)}$ are local coordinates on $\cU_\alpha$ and $\cU_\beta$, respectively. 
In particular, the patching condition for the $B$-field specified by the transformation matrix $\cF$ becomes
\begin{align}
 B_{(\alpha)mn} = \frac{\partial x_{(\beta)}^k}{\partial x_{(\alpha)}^m}\,\frac{\partial x_{(\beta)}^l}{\partial x_{(\alpha)}^n}\,\Bigl(B_{(\beta)kl}+\partial^{(\beta)}_{[k}\zeta_{(\alpha\beta) l]}\Bigr) + \partial^{(\alpha)}_{[m}\zeta_{(\alpha\beta)n]} \,.
\label{eq:B-patch-HZ}
\end{align}
This patching condition exhibits that, on a triple overlap of patches $\cU_\alpha\cap \cU_\beta \cap \cU_\gamma$, the transformation must satisfy the consistency condition:
\begin{align}
 \zeta_{(\alpha\beta)}+\zeta_{(\beta\gamma)}+\zeta_{(\gamma\alpha)} = 0 \,.
\label{eq:consistency-zeta}
\end{align}
From this consistency condition and \eqref{eq:B-patch-HZ}, Papadopoulos showed in \cite{Papadopoulos:2014mxa} that there exists a one-form, $\lambda_{(\alpha)}$, which can be used to find a globally defined $B$-field, $\widetilde{B}_{(\alpha)} \equiv B_{(\alpha)} + \rmd \lambda_{(\alpha)}$; $\widetilde{B}_{(\alpha)} = \widetilde{B}_{(\beta)}$ . 
This implies that the three-form field strength $H_3=\rmd B_2=\rmd \widetilde{B}_2$ is globally an exact form. 
This is so-called the 'Papadopoulos problem' \cite{Papadopoulos:2014mxa} that, if one requires the patching condition for the dual coordinates as in \eqref{eq:patch-generalized-coords}, one cannot describe general backgrounds with non-trivial $H$-flux. 

It is useful to contrast the situation to the generalized geometry \cite{Hitchin:2004ut,Gualtieri:2003dx} . There, the dual coordinates are not introduced, and the patching condition for coordinates and $B$-field are simply given by
\begin{align}
 x_{(\alpha)}^m = x_{(\alpha\beta)}(x_{(\beta)})\,,\quad 
 B_{(\alpha)} = B_{(\beta)}+\rmd \zeta_{(\alpha\beta)} \, . 
\end{align}
Most significantly, consistency at a triple overlap imposes the condition
\begin{align}
 \rmd \bigl(\zeta_{(\alpha\beta)}+\zeta_{(\beta\gamma)}+\zeta_{(\gamma\alpha)}\bigr) =0 \,,
\label{eq:consistency-GG}
\end{align}
which is weaker than \eqref{eq:consistency-zeta}, and this does not exclude a closed but not exact $H$-flux \cite{Papadopoulos:2014mxa}.
In this sense, it is better not to introduce the dual coordinates. 
In the same sense, it is better to formulate that  the dual coordinates are not changed, as in the conventional formulation of the supergravity or as in the approach of \cite{Hull:2014mxa}.

Here, we show that our proposed approach is not subject to the above issue, despite that the dual coordinates do transform under diffeomorphisms in our approach.%
\footnote{In \cite{Chaemjumrus:2015vap}, which appeared on the arXiv nearly two months after this paper was posted, it was commented that our approach treats the dual coordinates transformed and so also suffers from the Papadopoulos's issue as  in the approach of Hohm and Zwiebach. The following exposition reiterates, given our definition of the transformed dual coordinates, the reason why our approach do not have such an issue.} 
It is well known and explained below \eqref{eq:b_s-sol} that, because of the strong constraint, the doubled spacetime always admits trivial Killing vectors of the form, $V^M=\partial^Mf(x)$, which does not generate any gauge transformations. 
The presence of such trivial Killing vectors indicates that the doubled spacetime is different from the conventional $2d$-dimensional Riemannian manifold, as we cannot distinguish two points which are connected by a flow generated by the trivial Killing vector. 
More specifically, the points in the doubled spacetime is identified by the equivalence relation, $(x^m,\,\widetilde{x}_m)\simeq \bigl(x^m,\,\widetilde{x}_m+ \partial_m f(x)\bigr)$, which we call the trivial coordinate gauge symmetry, following \cite{Park:2013mpa}. 
A generalized vector, which induces a displacement in the doubled spacetime, is also defined up to an addition of a trivial Killing vector; $V^M=(v^m,\,\widetilde{v}_m)\simeq (v^m,\,\widetilde{v}_m+\partial_m f)$. 
Taking account of this equivalence relation, we are able to weaken the consistency condition \eqref{eq:consistency-zeta} to
\begin{align}
\boxed{~
 \zeta_{(\alpha\beta)}+\zeta_{(\beta\gamma)}+\zeta_{(\gamma\alpha)} = \rmd f_{(\alpha\beta\gamma)} \,,\quad
 \text{equivalently}, \quad
\rmd \bigl( \zeta_{(\alpha\beta)}+\zeta_{(\beta\gamma)}+\zeta_{(\gamma\alpha)}\bigr) = 0\,,~
 }
\end{align}
as equality for the dual components holds only up to an exact form. 
This is nothing but the same type of  consistency condition \eqref{eq:consistency-GG} as in the generalized geometry. 
As such, we see that there is no issue in allowing the transformations of dual coordinates in describing generic backgrounds  with non-trivial $H$-flux. 
Further, by considering an intersection of four patches, $\cU_\alpha\cap\cU_\beta\cap\cU_\gamma\cap\cU_\delta$, the quantization condition can be written as
\begin{align}
 f_{(\alpha\beta\gamma)} - f_{(\alpha\beta\delta)} + f_{(\alpha\gamma\delta)} - f_{(\beta\gamma\delta)} = 2\pi n\qquad (n\in \mathbb{Z}) \,.
\end{align}
In our approach, we treat the field $b_{mn}$ (which is independent of the trivial coordinate gauge symmetry and has a good geometric interpretation associated with embedding of the null subspace $\cN$ inside the doubled spacetime) as a more fundamental object than the dual coordinates, and hence prescribe the patching condition as
\begin{align}
 x^m_{(\alpha)}=x^m_{(\alpha\beta)}(x_{(\beta)}) \qquad \text{and} \qquad
 b_{(\alpha)mn} = b_{(\beta)mn} + 2\,\partial_{[m} \zeta_{(\alpha\beta) n]} \,. 
\label{eq:patching}
\end{align}

\begin{figure}[h]
\centering
\includegraphics[width=0.55\linewidth]{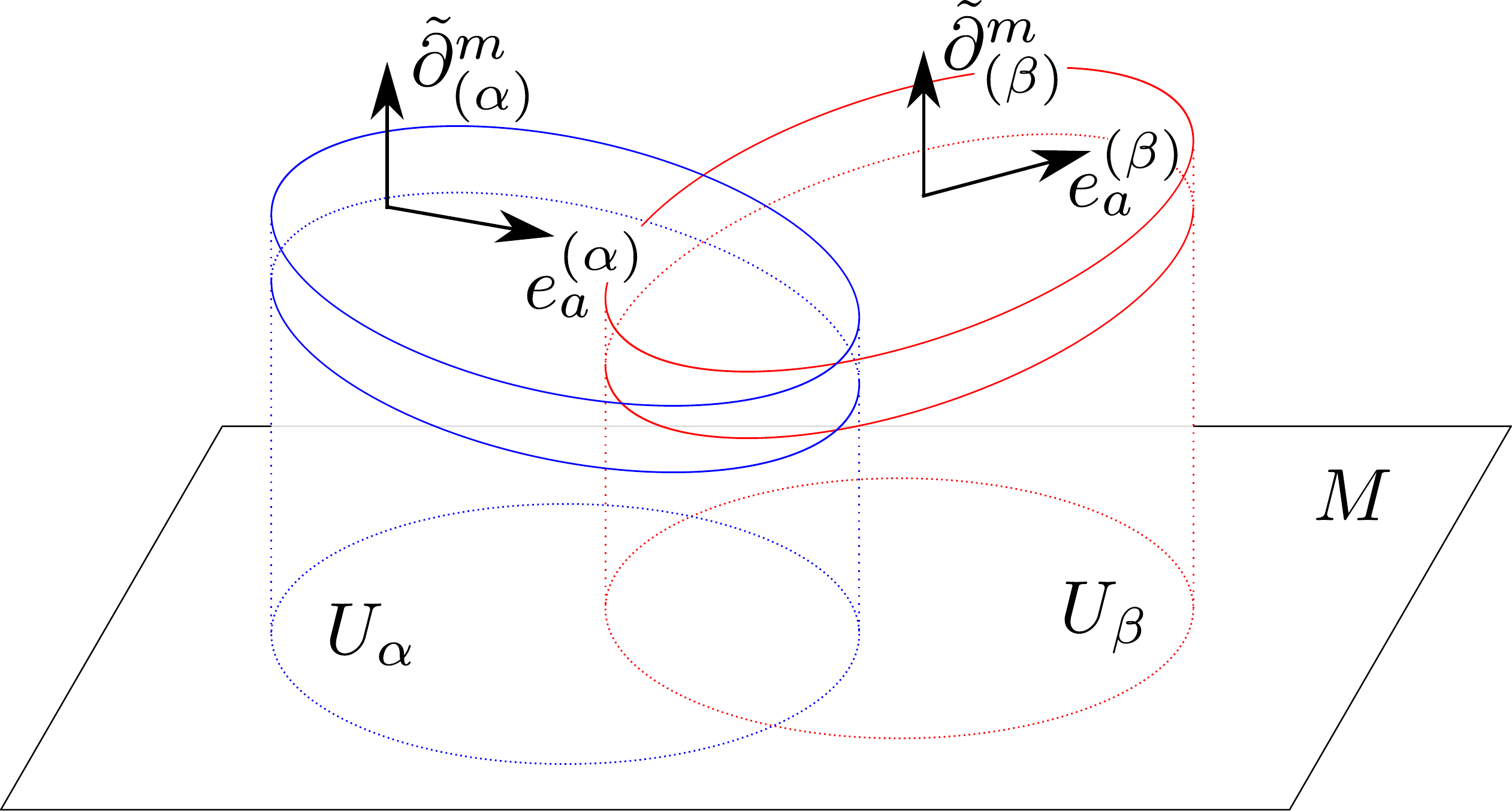}
\caption{Open sets $\cU_\alpha$ and $\cU_\beta$ (two `cylinders') with different foliations and the open sets $U_\alpha$ and $U_\beta$ on a common null plane $M$, a constant-$\tilde{x}_m$ plane. }
\label{fig:patch}
\end{figure}

Note that, since the closed two-form $b_{mn}(x)$ specifies the foliation of the doubled space by $d$-dimensional null surfaces, the patching condition \eqref{eq:patching} suggests that we are patching two open sets $\cU_\alpha$ and $\cU_\beta$ within different foliations (see Figure \ref{fig:patch}). 
As $b_{mn}(x)$ is independent of $\tilde{x}_m$, the shape of each leaf is uniform in all directions of dual coordinates $\tilde{x}_m$. 
The relation between the generalized tensors defined on the two patches is then given by \eqref{eq:our-formula} with the matrix $\ours^M{}_N$:
\begin{align}
 \ours^M{}_N =\begin{pmatrix} \delta^m_k & 0 \cr b^{(\alpha)}_{mk}(x_{(\alpha)}) & \delta_m^k \end{pmatrix}
\begin{pmatrix} \frac{\partial x_{(\alpha)}^k}{\partial x_{(\beta)}^l} & 0 \cr 0 & \frac{\partial x_{(\beta)}^l}{\partial x_{(\alpha)}^k} \end{pmatrix}\begin{pmatrix} \delta^l_n & 0 \cr -b^{(\beta)}_{ln}(x_{(\beta)}) & \delta_l^n \end{pmatrix} \,.
\end{align}
The rightmost matrix (or the inverse of leftmost matrix) maps a leaf of $\cU_\beta$ (or $\cU_\alpha$) to a subspace $U_\beta$ (or $U_\alpha$) on a common null plane $M$ of constant-$\tilde{x}_m$, while the middle matrix corresponds to a conventional diffeomorphism between $U_\alpha$ and $U_\beta$ (see Figure \ref{fig:patch}). 

Finally, we like to express the finite transformation for $b_{mn}$ in terms of the dual coordinates. 
We do this by first defining transformation law of the dual coordinates, $\tilde{x}^s_m(x)$. 
In our approach, it is given by the following flow equation:
\begin{align}
 \frac{\rmd}{\rmd s}\widetilde{x}^s_m(x) = \hat{\sfv}^{(s,V)}_m(x) \,.
\label{eq:diff-dual-coords}
\end{align}
This is different from the definition of Hohm and Zwiebach's definition, $\delta_V \widetilde{x}_m=\widetilde{v}_m$ (although, under a trivial coordinate gauge symmetry, our dual coordinates also transform just by an exact form).  The solution to \eqref{eq:diff-dual-coords} is formally obtained as
\begin{align}
 \widetilde{x}^s_m(x) = \widetilde{x}_m + \sum_{k=1}^\infty \frac{s^k}{k!} \bigl(\Lie_v^{k-1}\widehat{\sfv}\bigr)_m(x) 
 \equiv \widetilde{x}_m + \zeta^{(s,V)}_m(x) \,.
\end{align}
This solution is also expressible in the well-known form,
\begin{align}
 \widetilde{x}^s_m(x) 
 &= \widetilde{x}_m + \sum_{k=1}^\infty \frac{s^k}{k!} 
 \bigl[(v^n\partial_n + \rmd v^n\wedge \iota_n)^{k-1}\widehat{\sfv}\bigr]_m(x) 
\nn\\
 &= \Exp{s\sfV}\widetilde{x}_m + \sum_i \rho^s_i\partial_m\chi^s_i \, , 
\end{align}
where $\sfV^M\equiv (v^m,\,\widehat{\sfv}_m)$, $\rmd v^n\equiv \partial_k v^n\,\rmd x^k$, $\iota_n$ the interior product with $\partial_n$, and  $\rho^s_i,\chi^s_i$ functions of $v^m$, $\hat{\sfv}_m$ and $s$. 
In the simplest situation of $b_{mn}=0$, $\sfV^M=V^M$ and our transformed coordinates become $\widetilde{x}^s_m(x)\sim \Exp{sV}\widetilde{x}_m$ up to the coordinate gauge symmetry. 
Using the transformed dual coordinates as above, the finite transformation for $b_{mn}$ can now be expressed as
\begin{align}
\boxed{~
 b^{(s,V)}_{mn}(x) = b_{mn}(x) + \partial_m\widetilde{x}^s_n(x)-\partial_n\widetilde{x}^s_m(x) \,.~
 }
\end{align}
If we use the dual coordinates, the patching condition \eqref{eq:patching} can be also written as
\begin{align}
 x^m_{(\alpha)}=x^m_{(\alpha\beta)}(x_{(\beta)}) \qquad \text{and} \qquad
 \tilde{x}_{(\alpha)m} = \tilde{x}_{(\beta)m} + \zeta_{(\alpha\beta) m} \,,
\end{align}
although we should keep in mind the equivalence relation for the dual coordinates. 

Summarizing our approach, once we are given the gauge parameter, $V^M=(v^m,\,\widetilde{v}_m)$, and $b_{mn}(x)$, we can construct $b^{(s,V)}_{mn}$ or the pathlines, $x_s^M(x)=(\Exp{sv}x^m,\, \tilde{x}_m + \zeta^{(s,V)}_m)$, and, using it, we can express the transformed tensor as \eqref{eq:our-formula}. 
On the other hand, if $x_s^M(x)$ is given instead of the gauge parameter $V^M(x)$, the transformed tensor can be easily obtained by \eqref{eq:our-formula}. 
In this case, the corresponding gauge parameter can be read off as follows. 
First, we compute
\begin{align}
 \frac{\rmd}{\rmd s}x_s^M(x) = \bigl(v^m(x_s),\, \widehat{\sfv}^{(s,V)}_m(x)\bigr) \,,
\end{align}
and extract the first component, $v^m(x)$. 
Next, recalling $\widehat{\sfv}^{(s,V)}_m(x)= (\partial x_s^n/\partial x^m)\, \widehat{\sfv}_n(x_s)$, we can obtain $\widehat{\sfv}_m (= \widetilde{v}_m - b_{mn} \,v^n)$. 
As we already know $b_{mn}(x)$ and $v^m(x)$, we can find $\widetilde{v}_m(x)$. 
Finally, combining $v^m(x)$ and $\widetilde{v}_m(x)$, we find $V^M=(v^m,\,\widetilde{v}_m)$ completely. 

\subsection{Composition of finite transformations}
\label{sec:composition}
Here, we show that our approach exhibits composition property\footnote{See section 2.4 in \cite{Hohm:2013bwa} for the similar investigation, find the failure in the composition law for the transformation matrix $\cF^M{}_N$.}; 
two successive finite transformations can be consistently composed to a single transformation using our transformation matrix $\ours_M{}^N$. 
We start from the finite transformation law expressed in the push-forward form, 
\begin{align}
 W_{[s,V]}^M(x_s)
 &\equiv \ours^M{}_N (x_s, x) W^N(x)  \nonumber \\
 & = \begin{pmatrix} \delta^m_n & 0 \cr b^{(s,V)}_{mn}(x_s) & \delta_m^n \end{pmatrix}
\begin{pmatrix} \frac{\partial x_s^n}{\partial x^k} & 0 \cr 0 & \frac{\partial x^k}{\partial x_s^n} \end{pmatrix}
 \begin{pmatrix} \delta^k_l & 0 \cr -b_{kl}(x) & \delta_k^l \end{pmatrix}
 \begin{pmatrix} w^l(x) \cr \tilde{w}_l(x) \end{pmatrix} \,,
\end{align}
and consider a composition of two finite transformations, $\Exp{\gLie_{V_1}}$ followed by $\Exp{\gLie_{V_2}}$. 
Under the two successive transformations, we have
\begin{align}
 W_{(2;1)}^M(x_2)
 &= \begin{pmatrix} \delta^m_n & 0 \cr b^{(2;1)}_{mn}(x_2) & \delta_m^n \end{pmatrix}
\begin{pmatrix} \frac{\partial x_2^n}{\partial x_1^k} & 0 \cr 0 & \frac{\partial x_1^k}{\partial x_2^n} \end{pmatrix}
 \begin{pmatrix} \delta^k_l & 0 \cr -b^{(1)}_{kl}(x_1) & \delta_k^l \end{pmatrix}
\cdot
\nn\\
 &\quad\cdot
\begin{pmatrix} \delta^l_p & 0 \cr b^{(1)}_{lp}(x_1) & \delta_l^p \end{pmatrix}
\begin{pmatrix} \frac{\partial x_1^p}{\partial x^q} & 0 \cr 0 & \frac{\partial x^q}{\partial x_1^p} \end{pmatrix}
 \begin{pmatrix} \delta^q_r & 0 \cr -b_{qr}(x) & \delta_q^r \end{pmatrix}
 \begin{pmatrix} w^r(x) \cr \tilde{w}_r(x) \end{pmatrix} 
\nn\\
 &= \begin{pmatrix} \delta^m_n & 0 \cr b^{(2;1)}_{mn}(x_2) & \delta_m^n \end{pmatrix}
\begin{pmatrix} \frac{\partial x_2^n}{\partial x^k} & 0 \cr 0 & \frac{\partial x^k}{\partial x_2^n} \end{pmatrix}
 \begin{pmatrix} \delta^k_l & 0 \cr -b_{kl}(x) & \delta_k^l \end{pmatrix}
 \begin{pmatrix} w^l(x) \cr \tilde{w}_l(x) \end{pmatrix} \,.
\label{eq:finite-composition}
\end{align}
Here, we used the chain rule, and abbreviated $b^{(1)}_{mn}(x)\equiv b^{(s=1,V_1)}_{mn}(x)$ and
\begin{align}
 b^{(2;1)}_{mn}(x_1) 
 &\equiv b^{(1)}_{mn}(x_1) + 2\,\partial_{[m} \zeta^{(s=1,V_2;V_1)}_{n]}(x_1)\,,
\end{align}
where
\begin{align}
 \zeta^{(s,V_2;V_1)}_m(x_1)&\equiv \sum_{k=1}^\infty \frac{s^k}{k!}\,\bigl(\Lie_{v_2}^{k-1} \widehat{\sfv}^{(2;1)}\bigr)_m(x_1) \qquad \text{and} \qquad 
 \hat{\sfv}^{(2;1)}_m(x_1)\equiv (\widetilde{v}_2 +\iota_{v_2}b^{(1)})_m(x_1) \,.
\end{align}
The final form \eqref{eq:finite-composition} shows that the composition law is satisfied provided $b^{(2;1)}_{mn}(x_2)$ coincides with the transformed $b_{mn}$ under the combined single finite transformation, which we termed $b^{(2;1)}_{mn}(x)$. 

In order to construct the combined transformation, we need to consider diffeomorphisms in the doubled spacetime. We denote the transformed point under the first transformation, $\Exp{\gLie_{V_1}}$, by $x_1^M$, and under the second transformation, $\Exp{\gLie_{V_2}}$, by the point $x_2^M$ mapped from $x_1^M$. 
The dual components of the second orbit, $x_1^M\to x_2^M$, described in the original coordinates are given by
\begin{align}
 \widetilde{x}^s_m(x) 
 = \widetilde{x}^1_m(x) + \frac{\partial x_1^n}{\partial x^m}\,\zeta^{(s,V_2;V_1)}_n(x_1) 
 = \widetilde{x}_m + \bigl(\zeta^{(s=1;V_1)}_m + \zeta^{(s,V_2;V_1)}_m\bigr)(x) \,,
\end{align}
where we used that $\zeta^{(s,V_2;V_1)}_n(x)$ behaves as the conventional one-form. 
We obtain
\begin{align}
 \widetilde{x}_m^2(x) = \widetilde{x}_m + \zeta^{(21)}_m(x) \qquad \text{where} \qquad 
 \zeta^{(21)}_m(x)\equiv \zeta^{(s=1;V_1)}_m(x) + \zeta^{(s=1,V_2;V_1)}_m(x)\, . 
\end{align}
So, under the combined single transformation, $x^M\to x_2^M$, we obtain from our formula \eqref{eq:finite-bmn}, 
\begin{align}
 b^{(21)}_{mn}(x) = b_{mn}(x) + 2\,\partial_{[m}\zeta^{(21)}_{n]}(x) \,.
\end{align}
This indeed matches with the pullback of $b^{(2;1)}_{mn}(x_1)$ under the first diffeomorphism, $x^M\to x_1^M$:
\begin{align}
 b^{(2;1)}_{mn}(x) 
 & = b^{(1)}_{mn}(x) + 2\,\partial_{[m} \zeta^{(s=1,V_2;V_1)}_{n]}(x) \nonumber \\
& = b_{mn}(x) + 2\,\partial_{[m} \bigl(\zeta^{(s=1,V_1)}_{n]} +\zeta^{(s=1,V_2;V_1)}_{n]}\bigr)(x)\,.
\end{align}
From this, we conclude that the finite transformations in our approach satisfies the composition law. 
Note again that, although the dual coordinates are non-trivially transformed by $\delta \widetilde{x}^s_m$ or $\zeta^{(21)}_m$ is defined up to the addition of a locally exact form, their exterior derivatives are meaningful geometric quantities. 
Thus, the comparison between $b^{(2;1)}_{mn}(x)$ and $b^{(21)}_{mn}(x)$ is meaningful. 

Earlier, in our approach, we introduced the geometric quantity $b_{mn}(x)$ as a way recording rotation of the null surface inside the doubled spacetime. 
In this section, we found that the $b_{mn}$-dependence of the transformation law played an important role in satisfying the composition law, in particular, the cancellation of $b^{(1)}_{mn}(x_1)$ in \eqref{eq:finite-composition} was a crucial ingredient. 

\subsection{Finite transformations in non-geometric backgrounds}
\label{sec:non-geom}

We now extend consideration of our approach to non-geometric background. Again, we describe the background with local coordinates, $x^M=(x^m,\,\tilde{x}_m)$, where background fields are constrained to obey $\tilde{\partial}^m=0$. If the background admits abelian isometries, we can choose the conventional coordinates, $x^m=(x^a,\,x^i)$, such that the background fields are independent of $x^i$. In this case, the strong constraint is satisfied even if the gauge parameter $V^M$ depends on the dual coordinates $\widetilde{x}_i$ to the isometry directions since
\begin{align}
 \partial_N V^M(x)\,\partial^N \phi(x) 
 = \tilde{\partial}^i V^M(x) \,\partial_i \phi(x) = 0\,.
\end{align}
In this section, we explore this interesting possibility in detail. Of course, in order for the gauge algebra to be closed, gauge parameters $V^M$ cannot depend more than half of the coordinates. 

Let us first motivate why $\widetilde{x}_i$-dependent gauge transformation is physically interesting. 
Recall that, under the conventional constraint $\tilde{\partial}^m=0$ on gauge parameters, the gauge symmetry of DFT is a semi-direct product of diffeomorphisms and $B$-field transformations. 
On the other hand, the aforementioned possibility of the $\widetilde{x}_i$-dependence of gauge parameters provides different gauge symmetries, called $\beta$-transformations in place of $B$-field transformations. 
As discussed, for example, in \cite{Andriot:2011uh,Andriot:2012wx,Andriot:2012an,Hassler:2013wsa}, $\beta$-transformations play an important role in describing non-geometric backgrounds in DFT.  We thus learn that, in our approach, $\widetilde{x}_i$-dependent gauge transformations are relevant for describing non-geometric backgrounds. 

In what follows, when we consider a non-geometric background, we treat the gauge parameters as functions of $(x^a,\,\widetilde{x}_i)$. 
In this case, as discussed in \cite{Cederwall:2014opa}, fluctuations of the background fields should also be functions of $(x^a,\,\widetilde{x}_i)$ due to the strong constraint.%
\footnote{If we require the existence of isometries (i.e.~$\widetilde{\partial}^i=0$ for background fields) and also require that the background fields do not acquire the $\widetilde{x}_i$-dependence through the generalized diffeomorphisms, the non-isometric components, $v^a(x)$, should not have the $\widetilde{x}_i$-dependence and $v^i(x)$ can only have the linear dependence on $\tilde{x}_i$.} 
For simplicity, we consider only the extreme case where all fields are function only of $x^m$ or of $\tilde{x}_m$ (i.e.~no mixing between them). 

\subsubsection*{Non-geometric twisting}

The standard parameterization of the generalized metric in DFT is given by
\begin{align}
 \cH_{MN} = \begin{pmatrix} \delta_m^p & B_{mp} \cr 0 & \delta^m_p \end{pmatrix} 
\begin{pmatrix} G_{pq} & 0 \cr 0 & G^{pq} \end{pmatrix}
\begin{pmatrix} \delta^q_n & 0 \cr -B_{qn} & \delta_q^n \end{pmatrix} \, . 
\end{align}
This may be interpreted as a twisted form of the untwisted generalized metric \cite{Hull:2014mxa},
\begin{align}
 \hat{\cH}_{MN}\equiv \begin{pmatrix} G_{mn} & 0 \cr 0 & G^{mn} \end{pmatrix} \,.
\end{align}
On the other hand, in non-geometric backgrounds, the following non-geometric parameterization is known to be more convenient:
\begin{align}
 \cH_{MN} = \begin{pmatrix} \delta_m^p & 0 \cr \beta^{mp} & \delta^m_p \end{pmatrix} 
\begin{pmatrix} G_{pq} & 0 \cr 0 & G^{pq} \end{pmatrix}
\begin{pmatrix} \delta^q_n & -\beta^{qn} \cr 0 & \delta_q^n \end{pmatrix} \,.
\end{align}
This can be understood as providing an alternative twisting of $\hat{\cH}_{MN}$. 
In order to distinguish the two situations, we call the former the $H$-twisting while the latter the $R$-twisting. 

In order to exhibit the similarity between the $H$-twisting and $R$-twisting, we introduce a matrix, $L^M{}_N$, which is an $\OO(d,d)$ element, given by
\begin{align}
 L^M{}_N = \begin{pmatrix} \delta^m_n & 0 \cr B_{mn} & \delta_m^n \end{pmatrix}\qquad \text{or}\qquad
 L^M{}_N = \begin{pmatrix} \delta^m_n & \beta^{mn} \cr 0 & \delta_m^n \end{pmatrix}\,,
\label{eq:B-or-beta}
\end{align}
and express a generalized vector $V^M$ in terms of the untwisted vector, $\widehat{V}^M\equiv (\check{v}^m,\, \hat{v}_m)$, by
\begin{align}
 V^M \equiv L^M{}_N \, \widehat{V}^N \,. 
\end{align}
Then, the generalized Lie derivative can be rewritten as
\begin{align}
 \gLie_V W^M 
 &= L^M{}_N\, \bigl[ \widehat{V}^K D_K\widehat{W}^N -\widehat{W}^K \bigl(D_K\widehat{V}^N-D^N\widehat{V}_K\bigr)   + \widehat{V}^P\,\widehat{W}^Q\,\cF_{PQ}{}^N\bigr] 
\nn\\
 &= L^M{}_N\, \left[\begin{pmatrix}
 (\cL_{\check{v}} + \bar{\cL}_{\widehat{v}}) \check{w}^n
  +\hat{w}_k\,(\widetilde{D}^n \check{v}^k - \widetilde{D}^k \check{v}^n) \cr
 (\cL_{\check{v}} + \bar{\cL}_{\hat{v}}) \hat{w}_n 
 + \check{w}^k\,(D_n \hat{v}_k - D_k \hat{v}_n) 
\end{pmatrix}
  + \widehat{V}^P\,\widehat{W}^Q\,\cF_{PQ}{}^N\right] \,,
\label{eq:genLie-untwist}
\end{align}
where we defined $D_M \equiv (D_m,\,\tilde{D}^m) \equiv L^N{}_M\,\partial_N$ and
\begin{align}
\begin{split}
 \cL_{\check{v}}\hat{w}_m 
 &\equiv \check{v}^n\,D_n\hat{w}_m + \hat{w}_n\,D_m\check{v}^n \,,\qquad
 \cL_{\check{v}}\check{w}^m
 \equiv \check{v}^n\,D_n\check{w}^m - \check{w}^n\,D_n\check{v}^m \,,
\\
 \bar{\cL}_{\hat{v}}\check{w}^m
 &\equiv \hat{v}_n\,\widetilde{D}^n\check{w}^m + \check{w}^n\,\widetilde{D}^m\hat{v}_n \,,\qquad
 \bar{\cL}_{\hat{v}}\hat{w}_m 
 \equiv \hat{v}_n\,\widetilde{D}^n\hat{w}_m - \hat{w}_n\,\widetilde{D}^n\hat{v}_m \,,
\\
 \cF_{MNP} &\equiv 3\,\Omega_{[MNP]},\,\qquad 
 \Omega_{MNP} \equiv \eta_{RS}\,D_M L^R{}_N\,L^S{}_P = -\Omega_{MPN} \,.
\end{split}
\end{align}
Suppose we choose a parameterization for $L^M{}_N$ in terms of $B_{mn}$ (where $\check{v}^m=v^m$) and use the constraint $\widetilde{\partial}^m=0$. 
Then, the quantity inside the square bracket on the right-hand side of \eqref{eq:genLie-untwist} becomes
\begin{align}
 \begin{pmatrix}
 \Lie_{v} \,w^n \cr
 \Lie_{v} \,\hat{w}_n 
 + (\partial_n \hat{v}_k -\partial_k \hat{v}_n)\,w^k - v^k\,w^l\,H_{kln} 
\end{pmatrix} \,,\qquad H_{kmn}\equiv 3\,D_{[k}B_{mn]}= 3\,\partial_{[k}B_{mn]} \,.
\end{align}
We observe that this has the same form as the $H$-twisted Dorfman bracket \cite{Severa:2001qm}. 
Suppose we alternatively choose a parameterization in terms of $\beta^{mn}$ (where $\hat{v}_m=\tilde{v}_m$) and use the constraint $\partial_m=0$. 
Then, the same quantity becomes
\begin{align}
 \begin{pmatrix}
 \bar{\cL}_{\tilde{v}}\check{w}^n 
 + (\tilde{\partial}^n \check{v}^k -\tilde{\partial}^k \check{v}^n)\, \tilde{w}_k 
 - \tilde{v}_k\,\tilde{w}_l\,R^{kln} \cr
 \bar{\cL}_{\tilde{v}}\tilde{w}_n 
 \end{pmatrix} \,,\qquad R^{kmn}\equiv 3\,\widetilde{D}^{[k}\beta^{mn]} = 3\,\widetilde{\partial}^{[k} \beta^{mn]} \,.
\end{align}
This has the same form as the $R$-twisted Dorfman bracket discussed in \cite{Asakawa:2015jza} (if we identify $\tilde{\partial}^m$ with $\theta^{mn}\,\partial_n$ there, assuming $\theta^{mn}$ is constant). 

From $W^M = L^M{}_N\,\widehat{W}^N$, we can decompose the generalized Lie derivative as
\begin{align}
 \gLie_V W^M = L^M{}_N\,\delta_V \widehat{W}^N + \delta_V L^M{}_N\,\widehat{W}^N \,.
\end{align}
We find it natural to assume that the variation of the untwisted vector is given by
\begin{align}
 (\delta_V \widehat{W})^M = \begin{pmatrix}
 (\cL_{\check{v}} + \bar{\cL}_{\hat{v}}) \check{w}^m \cr
 (\cL_{\check{v}} + \bar{\cL}_{\hat{v}}) \hat{w}_m 
 \end{pmatrix} \, . 
\end{align}
For instance, this reduces to the conventional transformation \eqref{eq:finite-untwisted} in the case of the $H$-twisting with $\widetilde{\partial}^m=0$. 
Then, from \eqref{eq:genLie-untwist} and the relation $L^M{}_K\,\delta L^K{}_N=\delta L^M{}_N$ (which follows from \eqref{eq:B-or-beta}), the variation of $L^M{}_N$ should be given by
\begin{align}
 \delta_V L^M{}_N 
 = \begin{pmatrix} 0 & \tilde{D}^m \check{v}^n - \tilde{D}^n \check{v}^m \cr
 D_m \hat{v}_n - D_n \hat{v}_m & 0
\end{pmatrix} + \widehat{V}^K\, \cF_{KN}{}^M \,.
\end{align}
For either the $H$-twisting or the $R$-twisting, the transformation of the connection can be written as
\begin{align}
\begin{split}
 \delta_V B_{mn}&=\Lie_v B_{mn} + 2\,\partial_{[m}\tilde{v}_{n]} 
 = 2\,\partial_{[m}\hat{v}_{n]} + v^k\,H_{kmn} \,,
\\
 \delta_V \beta^{mn}&=\bar{\cL}_{\tilde{v}} \beta^{mn} + 2\,\widetilde{\partial}^{[m}v^{n]} 
 = 2\,\widetilde{\partial}^{[m}\check{v}^{n]} + \tilde{v}_k\,R^{kmn} \,. 
\end{split}
\end{align}
The second equation is equivalent to the anti-symmetric part of (2.39) in \cite{Hohm:2010jy}. 

As it is clear from the above expressions, the difference between the $H$-twisting and the $R$-twisting is only in the position of the indices. Hence, the finite transformation in non-geometric backgrounds can be obtained straightforwardly, as we explicitly show below. 

\subsubsection*{Finite transformation in non-geometric backgrounds}

As the structure is completely the same apart from the position of indices, following the same discussion as in section \ref{sec:Finite-geom}, we readily obtain the transformation matrix $\ours^M{}_N$ in non-geometric backgrounds as
\begin{align}
\begin{split}
 \ours^M{}_N &\equiv 
 \begin{pmatrix} \delta^m_k & \bar{\beta}_{(s,V)}^{mk}(x) \cr 0 & \delta_m^k \end{pmatrix}
 \begin{pmatrix} \frac{\partial\tilde{x}^s_l}{\partial \tilde{x}_k} & 0 \cr 0 & \frac{\partial\tilde{x}_k}{\partial \tilde{x}^s_l} \end{pmatrix}
 \begin{pmatrix} \delta^l_n & -\bar{\beta}^{ln}(x_s) \cr 0 & \delta_l^n \end{pmatrix}\,, 
\end{split}
\end{align}
where
\begin{align}
\begin{split}
  x_s^m(x) &\equiv x^m + \zeta_{(s,V)}^m(x) \,,\qquad 
 \widetilde{x}^s_m(x)= \Exp{s\,\tilde{v}_n\widetilde{\partial}^n}\tilde{x}_m \,,
\\
 \bar{\beta}_{(s,V)}^{mn}(x)&\equiv \bar{\beta}^{mn}(x) + 2\,\tilde{\partial}^{[m}\zeta_{(s,V)}^{n]}(x)\,,
\\
 \zeta_{(s,V)}^m(x)&\equiv \sum_{k=1}^\infty \frac{s^k}{k!}\,\bigl(\bar{\cL}_{\tilde{v}}^{k-1} \check{\sfv}\bigr)^m(x)
 = \int_0^s\rmd s'\, \check{\sfv}_{(s',V)}^m(x) \,,
\\
 \check{\sfv}_{(s,V)}^m(x) 
 &\equiv \Exp{s\bar{\cL}_{\tilde{v}}}\check{\sfv}^m(x) 
  = \frac{\partial\tilde{x}^s_n}{\partial \tilde{x}_m}\,\check{\sfv}^n(x_s)\,, \quad 
 \check{\sfv}^m(x)\equiv v^m(x)-\bar{\beta}^{mn}\,\tilde{v}_n(x) \,.
\end{split}
\end{align}
In this case, $\bar{\beta}^{ma}(x)$ is introduced as the upper components of a set of tangent vectors on a null surface $\cN$, on which the background fields and the gauge parameters are defined:
\begin{align}
 \widetilde{e}^{Ma}(x) = \begin{pmatrix} \bar{\beta}^{ma}(x) \cr \delta_m^a \end{pmatrix}\qquad \text{equivalently}\qquad
 \tilde{e}^a(x) = \widetilde{\partial}^a - \bar{\beta}^{am}(x) \, \partial_m \,.
\end{align}
Here as well, from null property $\eta_{MN}\,\tilde{e}^{Ma}\,\tilde{e}^{Nb}=0$ we find that $\bar{\beta}^{ma}(x)$ is anti-symmetric, and, from the gauge fixing $\tilde{e}_m{}^a(x) =\delta_m^a$, we obtain
\begin{align}
 \delta_V \bar{\beta}^{mn}(x) 
 = \bar{\cL}_{\tilde{v}}\bar{\beta}^{mn}(x) + 2\,\widetilde{\partial}^{[m}\tilde{v}^{n]}(x)\,.
\label{eq:beta-bar-gauge}
\end{align}
Further, requiring again that the generalized Lie derivative of a tangent vector on $\cN$ along an arbitrary tangent vector has only the tangential components, we obtain the closedness condition, $\widetilde{\partial}^{[k}\bar{\beta}^{mn]}=0$. 

We can continue using the same matrix $\ours^M{}_N$ even if there exist additional external directions, $x^\mu$. 
If we assume gauge parameters have the coordinate dependence $\bigl(\xi^\mu(x^\nu),\,V^M(\tilde{x}_m)\bigr)$ and the internal components of the background fields satisfy the condition $\partial_m=0$, the exponentiation of the generalized vector can be decomposed as
\begin{align}
 \bigl(\Exp{s(\Lie_\xi + \gLie_V)}W\bigr)^M
 = \Exp{s\,\xi} \bigl(\Exp{s\gLie_V}W\bigr)^M(x) 
 = \bigl(\Exp{s\gLie_V}W\bigr)^M(\Exp{s\,\xi}x) \,.
\end{align}
We see that the effect of the external gauge parameter $\xi^\mu$ shows up only through the coordinate dependence. 
In the examples considered in section \ref{sec:examples}, we indeed consider such cases with external directions, where $x^\mu=(t,r,\theta)$. 

\section{Examples}
\label{sec:examples}

In this section, we apply our proposal for the finite transformation law to two codimension-2 backgrounds -- one geometric background and another non-geometric background -- in DFT, and discuss some global aspects of these backgrounds. 

\subsubsection*{Defect NS5-brane background}

This geometric background is sourced by a NS5-brane, which is extended in the $x^3,\dotsc,x^7$ directions and smeared over the circle compactified $x^8, x^9$ directions of radii $R_8, R_9$:
\begin{align}
\begin{split}
 &\rmd s^2 = H(r)\, \bigl(\rmd r^2+r^2\,\rmd\theta^2\bigr) 
 + \rmd x^2_{03\cdots 7} + H(r)\,\rmd x_{89}^2 \,, \quad
 \Exp{2\phi} = H(r) \,,
\\
 &B_{2}= \frac{\sigma\, \theta}{2\pi}\,\rmd x^8\wedge \rmd x^9 \,,\qquad 
  H(r)\equiv \frac{\sigma}{2\pi}\,\log(r_\text{c}/r) \,.
\end{split}
\label{eq:NS5-BG}
\end{align}
Here, $r_\text{c}$ is a positive constant and $\sigma \equiv l_s^2/(R_8R_9)$. 
Since the NS5-brane is smeared along the $x^8$-$x^9$ torus, integrating the Gauss' law of the three-form flux $H_3 = \rmd B_2$ over the torus, the charge of NS5$(34567)$-brane  is given by
\begin{align}
 Q_{\text{NS5}}\equiv \frac{2\pi}{\sigma}\,\oint_C {\rmd B_{89} \over 2 \pi} \,,
\end{align}
where $C$ is a closed contour which encloses the NS5-brane on the $(r,\theta)$-plane. 

We now extend this background to a doubled geometry and introduce two (doubled) coordinate patches, $\cU_{\alpha}=\{\theta \in (-\pi/4,\,5\pi/4)\}$ and $\cU_{\beta}=\{\theta \in (3\pi/4,\,9\pi/4)\}$. 
Their overlap, $\cU_\alpha\cap \cU_\beta$, consists of $\cO_1=\{\theta \in (3\pi/4,\,5\pi/4)\}$ and $\cO_2=\{\theta \in (-\pi/4,\,\pi/4)=(7\pi/4,\,9\pi/4)\}$. 
We denote the $B$-field on the patch $\cU_\alpha$ as $B_{(\alpha)}=\frac{\sigma\, \theta_{(\alpha)}}{2\pi}\,\rmd x_{(\alpha)}^8\wedge \rmd x_{(\alpha)}^9$ and on the patch $\cU_\beta$ as $B_{(\beta)}=\frac{\sigma\, (\theta_{(\beta)}-2\pi)}{2\pi}\,\rmd x_{(\beta)}^8\wedge \rmd x_{(\beta)}^9$. 
On the overlap $\cO_1$, as the $B$-field is related by $B_{(\alpha)89}=B_{(\beta)89}+\sigma$, the patching condition is given by
\begin{align}
 x^m_{(\alpha)}=x^m_{(\beta)} \qquad \text{and} \qquad \quad b_{(\alpha)89} = b_{(\beta)89} + \sigma \,. 
\end{align}
In particular, we can choose $b_{(\alpha)89} =\sigma$ and $b_{(\beta)89} =0$\,. 
On the other hand, on the overlap $\cO_2$, the patching condition is given by
\begin{align}
 \theta_{(\alpha)}=\theta_{(\beta)}-2\pi\,,\quad x_{(\alpha)}^{(\text{others})}=x_{(\beta)}^{(\text{others})} \quad \text{and} \quad \quad b_{(\alpha)89} = b_{(\beta)89} + \sigma \,,
\label{eq:patch-killing}
\end{align}
which indeed gives $B_{(\alpha)89} = B_{(\beta)89}$. 
The charge of NS5-brane is given by
\begin{align}
 Q_{\text{NS5}} = \frac{1}{\sigma}\,\Big[B_{(\alpha)89}-B_{(\beta)89} \Bigr]^{\theta=\pi}_{\theta=0} = 1 \,.
\end{align}
As shown above, in the presence of the $H$-flux, we need to patch the doubled spacetime with different foliations. 
Namely, we cannot choose a global $d$-dimensional subspace with $b_{mn}=0$. 
The above NS5-brane background demonstrates this affirmatively. 

Let us now understand the patching condition \eqref{eq:patch-killing} from the DFT viewpoint. 
The doubled geometry of the smeared NS5-brane background actually has a generalized Killing vector $\xi^M$,
\begin{align}
 \xi \equiv 2\pi \partial_\theta - \sigma x^8\,\widetilde{\partial}^9 \,.
\end{align}
Therefore, we can identify the $\theta$ direction along the flow of $\xi^M$. 
If we calculate the pullback in our approach, we get $\zeta_9^{(s,\xi)}= -s \sigma x^8$ and $b^{(s,\xi)}_{89}(x) = b_{89}(x) - s \sigma$. So, we obtain the expected result,
\begin{align}
 \Exp{\gLie_\xi}\cH_{MN}(r,\theta) = \cH^{(s,\,\xi)}_{MN}(r,\theta) = \cH_{MN}(r,\theta) \,.
\end{align}
The angular coordinate $\theta$ is normalized with the period $2\pi$, so we identify the background with parameter $s=1$:
\begin{align}
 \bigl(\theta ,\,\tilde{x}_9,\,x^{(\text{others})}\bigr)
 \quad \sim \quad \bigl(\theta+2\pi ,\,\tilde{x}_9-\sigma \, x^8,\,x^{(\text{others})}\bigr) \,.
\label{eq:NS5-idenfity}
\end{align}
This finite transformation corresponds precisely to the patching condition \eqref{eq:patch-killing}. 

\subsubsection*{Exotic $5^2_2$-brane background}

As an example of non-geometric background, $T$-fold, consider the $5^2_2(3\cdots 7,89)$-brane background \cite{LozanoTellechea:2000mc,deBoer:2010ud,deBoer:2012ma}, which can be obtained from \eqref{eq:NS5-BG} by performing a double $T$-duality in the directions of $x^8, x^9$-torus:
\begin{align}
\begin{split}
 &\rmd s^2 = H(r)\, \rmd x_{12}^2 + \rmd x^2_{03\cdots 7} + \frac{H(r)}{K(r,\,\theta)}\, \rmd x_{89}^2 \,, \quad
 \Exp{2\phi} = \frac{H(r)}{K(r,\,\theta)}\,,
\\
 &B_{2} = -\frac{\sigma\, \theta}{2\pi\,K(r,\,\theta)}\,\rmd x^8\wedge \rmd x^9 \,,\qquad 
 K(r,\,\theta) \equiv H^2(r) + \frac{\sigma^2\, \theta^2}{(2\pi)^2} \,.
\end{split}
\end{align}
Here, we regard $x^\mu=(t,r,\theta)$ are the external directions and all other directions as compactified toroidal directions. The harmonic function $H(r)$ is the same as \eqref{eq:NS5-BG} except that $\sigma$ here is given by $\sigma=R_8R_9/l_s^2$.  In the non-geometric description, the background is given by
\begin{align}
 \rmd s^2 = H(r)\, \rmd x_{12}^2 + \rmd x^2_{03\cdots 7} + H^{-1}(r)\, \rmd x_{89}^2 \,, \qquad
 \Exp{2\widetilde{\phi}} =  H^{-1}(r) \,, \qquad 
 \beta^{89} = \frac{\sigma\, \theta}{2\pi} \,,
\end{align}
where $\widetilde{\phi}$ is defined via $\Exp{-2\widetilde{\phi}} \sqrt{-\tilde{G}}= \Exp{-2\phi}\sqrt{-G}$\,. 

Suppose (similar to the situation in \cite{Kikuchi:2012za}) we allow the gauge parameter to depend on the dual coordinates $\widetilde{x}_m$. We then find that this background also admits a generalized Killing vector:
\begin{align}
 \xi_{5^2_2} = 2\pi \partial_\theta - \sigma\, \widetilde{x}_8\,\partial_9 \,.
\end{align}
Under a finite transformation along this generalized Killing vector, we have $\zeta_{(s,\xi)}^9(x) = - \sigma s\, \tilde{x}_8$, so the transformed coordinates are given by
\begin{align}
 (\theta_s,\,x_s^9,\,x_s^{(\text{others})})=\bigl(\theta+2\pi s,\,x^9-\sigma s\, \widetilde{x}_8,\,x^{(\text{others})}\bigr)\,.
\end{align}
This leaves the generalized metric invariant, $\cH^{(s,\,\xi_{5^2_2})}_{MN}(x) = \cH_{MN}(x)$. So, we identify the physical points with $s=1$;
\begin{align}
 \bigl(\theta ,\,x^9,\,x^{(\text{others})}\bigr)
 \sim \bigl(\theta+2\pi ,\,x^9-\sigma\, \tilde{x}_8,\,x^{(\text{others})}\bigr) \,.
\label{eq:522-identify}
\end{align}

Using \eqref{eq:522-identify}, one can consider the patching condition in the manner same as the NS5-brane background. We introduce the patches as in the case of the NS5-brane background and define the $\beta$-field on the patch $\cU_\alpha$ as $\beta^{89}_{(\alpha)}=\frac{\sigma\, \theta_{(\alpha)}}{2\pi}$ and on the patch $\cU_\beta$ as $\beta^{89}_{(\beta)}=\frac{\sigma\, (\theta_{(\beta)}-2\pi)}{2\pi}$. 
We then find that the patching condition on $\cO_1$ is given by
\begin{align}
 x^m_{(\alpha)}=x^m_{(\beta)} \,,\quad \bar{\beta}_{(\alpha)}^{89} = \bar{\beta}_{(\beta)}^{89} + \sigma \,. 
\end{align}
while, on $\cO_2$, it becomes
\begin{align}
 \theta_{(\alpha)}=\theta_{(\beta)}-2\pi\,,\qquad x_{(\alpha)}^{(\text{others})}=x_{(\beta)}^{(\text{others})} \,,\qquad \beta_{(\alpha)}^{89} = \beta_{(\beta)}^{89} + \sigma \,.
\end{align}
The conserved charge of the smeared $5^2_2$-brane is given by
\begin{align}
 Q_{5^2_2}\equiv \frac{2\pi}{\sigma}\,\oint_C {\rmd \beta^{89} \over 2 \pi} = \frac{1}{\sigma}\,\Bigl[\beta_{(\alpha)}^{89}-\beta_{(\beta)}^{89} \Bigr]^{\theta=\pi}_{\theta=0} \, = 1 \,.
\end{align}

\vskip0.2cm
\section{Finite transformations in exceptional field theory}
\label{sec:finite-SL5}
The key feature of our approach is the use of untwisted vector for the doubled spacetime. 
The advantage of this feature is that it is straightforwardly generalizable to the EFTs. 
In this section, we demonstrate this by explicitly working out the simplest situation, the SL(5) EFT. 
Its generalization for bigger exceptional groups is straightforward and will be left as possible research projects for curious readers. 

\subsection{Review of SL(5) EFT}
We begin with a brief review of the effective theory for the M-theory compactified on a four-torus $\mathbb{T}^4$. 
We introduce the coordinates, $x^\mu$ ($\mu=0,\dotsc,6$), for the non-compact directions, and $x^i$ ($i=7,8,9,10$) for the compactified directions. 
Here, we are interested only in the compactified internal space $\mathbb{T}^4$. 
The relevant part of the Lagrangian density (i.e.~the potential involving variations over the internal space) is given by
\begin{align}
 \cL_{\text{internal}} = R(G)-\frac{1}{2\cdot 4!}\, F_{i_1\cdots i_4}\,F^{i_1\cdots i_4}\,,
\end{align}
where $R(G)$ is the Ricci scalar for the internal metric $G_{ij}(x)$ and $F_{i_1\cdots i_4}(x)\equiv 4\,\partial_{[i_1} C_{i_2i_3i_4]}(x)$ is the field strength associated with the three-form potential, $C_{ijk}(x)$. 
The indices are raised or lowered by using the internal metric $G_{ij}(x)$. 

This theory has two gauge symmetries, the conventional diffeomorphism and the gauge symmetry of the three-form potential. 
The gauge parameter for the internal diffeomorphism is a vector field $v(x)\equiv v^i(x)\,\partial_i$ and that for the internal three-form potential is a conventional two-form field defined on $\mathbb{T}^4$: $\hat{v}(x)\equiv (1/2)\,\hat{v}_{ij}(x)\,\rmd x^i\wedge\rmd x^j$. 
In \cite{Berman:2010is}, on the basis of the canonical formulation of the eleven-dimensional supergravity, the gauge algebra was examined and the gauge transformation for the metric and the three-form potential were obtained as
\begin{align}
 \delta_V G_{ij}(x)=\Lie_v G_{ij}(x)\,\quad \text{and} \quad 
 \delta_V C_{ijk}(x)=3\,\partial_{[i}\hat{v}_{jk]}(x) + v^l\,F_{lijk}(x) \,,
\label{eq:gauge-EFT}
\end{align}
where $\Lie_v$ is the conventional Lie derivative. 
Since a conventional vector field $w^i(x)$ and a two-form field $\hat{w}_{ij}(x)$ should be invariant under the gauge transformation for the three-form potential, if we combine them as an untwisted generalized vector,
\begin{align}
 \widehat{W}^M(x) \equiv \begin{pmatrix}
 w^i(x) \cr \frac{1}{\sqrt{2}}\hat{w}_{i_1i_2}(x) 
 \end{pmatrix} \,,
\end{align}
its transformation should be given by
\begin{align}
 \delta_V \widehat{W}^M(x) = \begin{pmatrix}
 \Lie_v w^i(x) \cr \frac{1}{\sqrt{2}}\Lie_v\hat{w}_{i_1i_2}(x) 
 \end{pmatrix} \,.
\end{align}
Now, using the three-form potential, we define a (twisted) generalized vector $W^M(x)$ by (along the lines of  \cite{Hull:2007zu,Berman:2010is})
\begin{align}
 W^M(x) \equiv \begin{pmatrix}
 w^i(x) \cr \frac{1}{\sqrt{2}}\,\tilde{w}_{i_1i_2}(x) 
 \end{pmatrix}
 \equiv \begin{pmatrix}
 \delta^i_j & 0 \\
 -\frac{1}{\sqrt{2}}\, C_{i_1i_2j}(x) & \delta^{j_1j_2}_{i_1i_2} \end{pmatrix} \widehat{W}^N(x) \,,
\end{align}
which can be shown to transform as
\begin{align}
 \delta_V W^M(x) = 
\begin{pmatrix}
 \Lie_v w^i(x) \cr
 \frac{1}{\sqrt{2}}\,\bigl[\Lie_v\tilde{w}_{i_1i_2}(x)-3\,\partial_{[i_1}\tilde{v}_{i_2j]}(x)\,w^j(x)\bigr]
\end{pmatrix} \,.
\label{eq:gen-Lie-section}
\end{align}
If we rewrite the lower component as
\begin{align}
 \delta_V \tilde{w}(x) = \Lie_v \tilde{w}(x) - \iota_w \rmd \tilde{v}(x) \,,
\end{align}
we can identify \eqref{eq:gen-Lie-section} with the Dorfman bracket \cite{Berman:2011cg},
\begin{align}
 &[v+\tilde{v},\,w+\tilde{w}]_D 
  \equiv [v,\,w] + \Lie_v \tilde{w} - \iota_w \rmd \tilde{v} \,.
\end{align}
Note that, in terms of the generalized vector field, variations of the metric and the three-form potential \eqref{eq:gauge-EFT} become the well-known form:
\begin{align}
 \delta_V G_{ij}(x)=\Lie_v G_{ij}(x)\,,\quad 
 \delta_V C_{ijk}(x)=\Lie_v C_{ijk}(x) + 3\,\partial_{[i}\tilde{v}_{jk]}(x) \,.
\end{align}

The M-theory compactified on a four-torus $\mathbb{T}^4$ is known to have the $\SL(5)$ $U$-duality symmetry. 
From the worldvolume theory of an M2-brane, DFT-like extension of the effective theory, the $\SL(5)$ EFT, was proposed in \cite{Berman:2010is}. 
It was further developed in \cite{Berman:2011cg}. 
In the $\SL(5)$ EFT, one introduces additional six coordinates, so-called winding coordinates, $y_{ij}\,(=y_{[ij]})$, which encode all possible winding configurations of the membrane.
Here, we omit the geometry of the external space, and only highlight the geometry of the ten-dimensional internal space with coordinates, $x^M=(x^i,\, y_{ij})$ ($M=1,\dotsc,10$). 
The geometry of the internal space is described by the generalized metric, which can be parameterized as%
\footnote{More generally, in order to make the generalized metric globally well-defined in non-geometric backgrounds, it is necessary to multiply a certain weight factor in front of the matrix.}
\begin{align}
 \cM_{MN} &= \begin{pmatrix}
 \delta_i^k & \frac{1}{\sqrt{2}}\, C_{ik_1k_2} \\
 0 & \delta^{i_1i_2}_{k_1k_2} \end{pmatrix}
\begin{pmatrix}
 G_{kl} & 0 \\
 0 & G^{k_1k_2,l_1l_2} \end{pmatrix}
\begin{pmatrix}
 \delta^l_j & 0 \\
 \frac{1}{\sqrt{2}}\, C_{l_1l_2j} & \delta_{l_1l_2}^{j_1j_2} \end{pmatrix}
\nn\\
 &= \begin{pmatrix}
 G_{ij} +\frac{1}{2}\,C_{ikl}\,C^{kl}{}_j & \frac{1}{\sqrt{2}}\, C_i{}^{j_1j_2} \\
 \frac{1}{\sqrt{2}}\, C^{i_1i_2}{}_j & G^{i_1i_2,j_1j_2} \end{pmatrix}\,,
\end{align}
where we defined $G^{i_1i_2,j_1j_2}\equiv G^{i_1[j_1}G^{j_2]i_2}$ 
and $\delta_{l_1l_2}^{j_1j_2}\equiv \delta_{[l_1}^{[j_1}\delta_{l_2]}^{j_2]}$. 
In order to make the $\SL(5)$ symmetry manifest, we also introduce new coordinates transforming in the ten-dimensional representation of the $\SL(5)$:
\begin{align}
 x^{ab} = \left\{\begin{array}{l}
 x^{i5} = x^i = - x^{5i} \cr x^{ij}=\frac{1}{2}\,\eta^{ijkl}\,y_{kl}
 \end{array}\right. \qquad \quad (a=1,\dotsc,5)\,,
\end{align}
where $\eta^{ijkl}$ (or $\eta_{ijkl}$) is a totally antisymmetric Levi-Civita symbol with the orientation convention $\eta^{1234}=1$ (or $\eta_{1234}=1$). 
In these new coordinates, the generalized metric is parameterized as \cite{Berman:2011cg}
\begin{align}
 \cM_{[a_1a_2][b_1b_2]} &= \begin{pmatrix}
 G_{ij} +\frac{1}{2}\,C_{ikl}\,C^{kl}{}_j & \frac{1}{2\sqrt{2}}\, C_i{}^{k_1k_2}\,\eta_{k_1k_2 j_1j_2} \\
 \frac{1}{2\sqrt{2}}\,\eta_{i_1i_2 k_1k_2}\, C^{k_1k_2}{}_j & \frac{1}{\det G_{ij}}\,G_{i_1i_2,j_1j_2} \end{pmatrix}\,.
\end{align}
Further, using the Hodge-dual variables, $\tilde{v}^{ij}\equiv \frac{1}{2}\,\eta^{ijkl}\,\tilde{v}_{kl}$ and $\tilde{w}^{ij}\equiv \frac{1}{2}\,\eta^{ijkl}\,\tilde{w}_{kl}$, the gauge transformation \eqref{eq:gen-Lie-section} can be rewritten as
\begin{align}
 \delta_V W^{a_1a_2} = \begin{pmatrix}
  \Lie_v w^i \cr
  v^k\,\partial_k \tilde{w}^{i_1i_2} 
 + 2\,\tilde{w}^{k[i_1}\,\partial_k v^{i_2]} 
 + \tilde{w}^{i_1i_2}\,\partial_k v^k
 + 2\,w^{[i_1}\,\partial_k \tilde{v}^{i_2]k}
\end{pmatrix} \,.
\end{align}
Upon imposing the constraint for the $\SL(5)$ EFT, $\partial_{ij}\equiv (\partial/\partial x^{ij}) =0$, we can generalize the gauge transformation to the generalized Lie derivative:
\begin{align}
 &\bigl(\gLie_V W\bigr)^{a_1a_2} \equiv \frac{1}{2}\,V^{cd}\,\partial_{cd} W^{a_1a_2} 
 + \frac{1}{2}\,W^{a_1a_2}\,\partial_{cd}V^{cd}
 + W^{a_1c}\,\partial_{cd}V^{da_2}
 -W^{ca_2}\,\partial_{cd}V^{a_1d} 
\nn\\
 &= \frac{1}{2}\,V^{b_1b_2}\,\partial_{b_1b_2} W^{a_1a_2} 
  - \frac{1}{2}\,W^{b_1b_2}\,\partial_{b_1b_2} V^{a_1a_2} 
  + \frac{1}{8}\,\varepsilon^{ea_1a_2b_1b_2}\,\varepsilon_{ec_1c_2d_1d_2}\,\partial_{b_1b_2}V^{c_1c_2}\,W^{d_1d_2} \,,
\end{align}
where $\varepsilon^{abcde}$ and $\varepsilon_{abcde}$ are another totally antisymmetric Levi-Civita symbols with $\varepsilon^{12345}=1=\varepsilon_{12345}$.
If we further introduce a notation, $A=[ab]$ ($a<b$), this can be written in a simpler form \cite{Berman:2012vc};
\begin{align}
 \gLie_V W^A &= V^B\,\partial_B W^A - W^B\,\partial_B V^A 
  + \varepsilon^{eAB}\,\varepsilon_{eCD}\,\partial_B V^C\,W^D \,.
\end{align}
The closure of the gauge algebra requires the strong constraint, $\varepsilon^{eAB}\,\partial_A\circ\,\partial_B\,\circ =0$, for arbitrary physical fields and gauge parameters. 
In this paper, we consider only a solution of the constraint, $\partial_{ij}=0$.\footnote{Another solution, called the IIB section, was found in \cite{Blair:2013gqa}. }

The Lagrangian (or the potential) of the $\SL(5)$ EFT was constructed in \cite{Berman:2010is}:
\begin{align}
 \cL &= \frac{1}{12}\,\cM^{MN}\,\partial_M\cM^{KL}\,\partial_N \cM_{KL} 
  -\frac{1}{2}\,\cM^{MN}\,\partial_N \cM^{KL}\,\partial_L \cM_{MK}
\nn\\
 &\quad +\frac{1}{12}\,\cM^{MN}\,\bigl(\cM^{KL}\, \partial_M \cM_{KL}\bigr)\,\bigl(\cM^{RS}\,\partial_N \cM_{RS} \bigr) 
\nn\\
 &\quad +\frac{1}{4}\,\cM^{MN}\,\cM^{PQ}\,\bigl(\cM^{RS}\, \partial_P \cM_{RS}\bigr)\,\bigl(\partial_M \cM_{NQ} \bigr)\,,
\end{align}
The corresponding internal action is invariant under the generalized diffeomorphism. 
See also \cite{Park:2013gaj} for the construction of the same action (up to surface integral) based on the differential geometry in the extended space, $U$-geometry. 

\subsection{Finite transformation law}

We now show how to construct the finite transformation law in SL(5) EFT following our proposed approach presented in section \ref{sec:Finite-geom}. 
For simplicity, we use the coordinates, $x^M=(x^i,\, y_{ij})$, although the final result can also be expressed in the $x^A$ coordinates. The finite transformation law for the untwisted vector fields is readily obtained as
\begin{align}
 \widehat{W}_{(s,V)}^M(x) = \begin{pmatrix} \frac{\partial x^i}{\partial x_s^j} & 0 \cr
 0 & \frac{\partial x_s^{[j_1}}{\partial x^{[i_1}}\,\frac{\partial x_s^{j_2]}}{\partial x^{i_2]}}
 \end{pmatrix}\begin{pmatrix} w^j(x_s) \cr \frac{1}{\sqrt{2}}\,\hat{w}_{j_1j_2}(x_s)
 \end{pmatrix}\,.
\end{align}
On the other hand, in order to obtain the transformation law for the three-form potential $C_{ijk}(x)$, we first introduce a gerbe connection $c_{ijk}(x)$ that satisfies 
\begin{align}
c_{ijk}(x) = c_{[ijk]}(x)\,, \qquad \partial_{[i} c_{jkl]}(x) = 0\,, \qquad \text{and} \qquad
 \delta_V c_{ijk}(x) = \Lie_v c_{ijk}(x) + 3\,\partial_{[i} \tilde{v}_{jk]}(x)  \,.
\label{prop}
\end{align}
This is done as follows. 
As in the case of DFT, we identify the dual component of tangent vectors on a four-dimensional null subspace $\cN$ (on which the physical fields are defined) as the gerbe connection:
\begin{align}
 e^M{}_\alpha(x) =\begin{pmatrix} \delta_\alpha^i\cr -\frac{1}{\sqrt{2}}\,c_{ij\alpha}(x)\end{pmatrix} \qquad \text{where} \qquad c_{ij\alpha}=c_{[ij]\alpha}\,.
\end{align}
It then satisfies the requisite properties \eqref{prop}. 
Firstly, it is partially symmetric, $c_{ij\alpha}=c_{\alpha ij}$, and this follows from the condition $(\iota_{v_{(1)}} \mu^{(2)}+\iota_{v_{(2)}} \mu^{(1)})_i=0$%
\footnote{In the SL(5)-covariant coordinates, $x^A$, this condition corresponds to the null property \eqref{eq:DFT-null} in DFT; $\epsilon_{eAB}\,V_{(1)}^A\,V_{(2)}^B=0$ (the fifth component, $\epsilon_{5AB}\,V_{(1)}^A\,V_{(2)}^B=0$, is trivially satisfied). This is required from the section condition in SL(5) EFT.}
for any tangent vectors $V_{(n)}^M=\bigl(v_{(n)}^i,\, \frac{1}{\sqrt{2}}\,\mu^{(n)}_{ij}\bigr)$ ($n=1,2$) with $\mu^{(n)}_{ij}\equiv -c_{ijk}\,v_{(n)}^k$. 
Secondly, it is closed, $\partial_{[i} c_{jkl]} =0$, and this follows from the requirement that the generalized Lie derivative of a tangent vector, $W^M=e^M{}_\alpha\, w^\alpha$, along a tangent vector, $V^M=e^M{}_\alpha\, v^\alpha$, 
\begin{align}
 \gLie_V W^M = \begin{pmatrix}
 \Lie_v w^i \cr
 \frac{1}{\sqrt{2}}\, \bigl(-c_{i_1i_2 k}\,\Lie_v w^k + 4\,\partial_{[k}c_{i_1i_2l]}\,v^k w^l \bigr)
\end{pmatrix} \,,
\end{align}
is again expanded by the tangent vectors, $e^M{}_\alpha$\,. 
Lastly, the variation $\delta_V c_{ijk}(x)$ is given as in \eqref{prop}, and this follows from the gauge condition, $e^i{}_\alpha=\delta^i_\alpha$.

Now, the finite transformation can be obtained by solving the differential equation:
\begin{align}
 \frac{\rmd}{\rmd s}c^{(s,V)}_{ijk}(x) 
 = \Lie_v c^{(s,V)}_{ijk}(x) + 3\,\partial_{[i}\tilde{v}_{jk]}(x) \,.
\end{align}
We can express the solution by introducing a two-form,
\begin{align}
 \hat{\sfv}_{ij}(x) \equiv \tilde{v}_{ij}(x) + c_{ijk}(x)\,v^k(x) \,,
\end{align}
such that
\begin{align}
 c^{(s,V)}_{ijk}(x) = c_{ijk}(x) + \sum_{n=1}^\infty \frac{s^n}{n!}\,\bigl(\Lie_v^{n-1}\rmd \hat{\sfv}\bigr)_{ijk} 
 \equiv c_{ijk}(x) + 3\,\partial_{[i} \zeta^{(s,V)}_{jk]}(x) \, . 
\label{eq:c-finite}
\end{align}
Here, 
\begin{align}
\begin{split}
 \zeta^{(s,V)}_{ij}(x)&\equiv \sum_{n=1}^\infty \frac{s^n}{n!}\,\bigl(\Lie_v^{n-1} \hat{\sfv}\bigr)_{ij}(x)
 = \int_0^s\rmd s'\, \hat{\sfv}^{(s',V)}_{ij}(x) \,,
\\
 \hat{\sfv}^{(s,V)}_{ij}(x) 
 &\equiv \Exp{s\Lie_v}\hat{\sfv}_{ij}(x) = \frac{\partial x_s^k}{\partial x^i}\,\frac{\partial x_s^l}{\partial x^j}\, \hat{\sfv}_{kl}(x_s) \, , 
\end{split}
\end{align}
where $x_s^m\equiv \Exp{s\,v}x^m$. 
In SL(5) EFT, there is a trivial Killing vector has the form, $V^M=\bigl(0,\, \partial_{[i}\alpha_{j]}\bigr)$.%
\footnote{In the SL(5)-covariant coordinates $x^A$, this can be written as 
$V^A(x)=- \varepsilon^{ABc}\partial_B\alpha_c(x)$, which indeed satisfies $\gLie_V W^A=0$ for an arbitrary vector field $W^A(x)$.} 
The corresponding $\zeta^{(s,V)}_{ij}(x)$ becomes locally an exact two-form, so it does not generate a gauge transformation. 
Therefore, as in the case of DFT, we identify two gauge parameters related by a trivial Killing vector as the same gauge parameter. 

Now, as in section 3 for DFT,  one can show that the combination $\bC_{ijk}(x)\equiv C_{ijk}(x)-c_{ijk}(x)$ transforms as
\begin{align}
 \delta_V \bC_{ijk}(x) = \Lie_v \bC_{ijk}(x) \,,
\end{align}
Its finite transformation is simply given by
\begin{align}
 \bC^{(s,V)}_{i_1i_2i_3}(x)
 = \frac{\partial x_s^{[j_1}}{\partial x^{[i_1}}\,\frac{\partial x_s^{j_2}}{\partial x^{i_2}}\,\frac{\partial x_s^{j_3]}}{\partial x^{i_3]}}\,
 \bC_{j_1j_2j_3}(x_s) \,.
\end{align}
Combining this with \eqref{eq:c-finite}, we obtain
\begin{align}
\boxed{~
 C^{(s,V)}_{i_1i_2i_3}(x) = \frac{\partial x_s^{[j_1}}{\partial x^{[i_1}}\,\frac{\partial x_s^{j_2}}{\partial x^{i_2}}\,\frac{\partial x_s^{j_3]}}{\partial x^{i_3]}}\,
 \bigl(C_{j_1j_2j_3} -c_{j_1j_2j_3}\bigr)(x_s) 
  + c_{i_1i_2i_3}(x) + 3\,\partial_{[i_1}\zeta^{(s,V)}_{i_2i_3]}(x) \,.~
}
\end{align}
We emphasize again that the finite transformation depends not only gauge parameters but also local specification of the null surface $\cN$ by $c_{ijk}(x)$, which can be chosen arbitrarily.

For a generalized vector, $W^M(x)$, we have the finite transformation
\begin{align}
 W_{(s,V)}^M(x) 
 &= \begin{pmatrix}
 \delta^i_j & 0 \\
 -\frac{1}{\sqrt{2}}\, C^{(s,V)}_{i_1i_2j}(x) & \delta^{j_1j_2}_{i_1i_2} \end{pmatrix} \begin{pmatrix} w_{(s,V)}^j(x) \cr \frac{1}{\sqrt{2}}\, \hat{w}^{(s,V)}_{j_1j_2}(x)
 \end{pmatrix}
 = \ours^M{}_N\,W^N(x_s) 
\end{align}
with
\begin{align}
 \ours^M{}_N\equiv \begin{pmatrix}
 \delta^i_k & 0 \\
 -\frac{1}{\sqrt{2}}\, c^{(s,V)}_{i_1i_2 k}(x) & \delta^{k_1k_2}_{i_1i_2} \end{pmatrix} 
 \begin{pmatrix}
 \frac{\partial x^k}{\partial x_s^l} & 0 \\
 0 & \frac{\partial x_s^{[l_1}}{\partial x^{[k_1}}\,\frac{\partial x_s^{l_2]}}{\partial x^{k_2]}} \end{pmatrix} 
\begin{pmatrix}
 \delta^l_j & 0 \\
 \frac{1}{\sqrt{2}}\, c_{l_1l_2j}(x_s) & \delta^{j_1j_2}_{l_1l_2} \end{pmatrix}\,.
\end{align}
We see that each matrix, and hence $\ours^M{}_N$ as well, is an element of $\SL(5)$ duality symmetry group.%
\footnote{See \cite{Berman:2011jh,Malek:2012pw} for the matrix representation of the exceptional group elements.} 

We close EFT consideration with a comment.  We can alternatively define the transformed dual coordinates, $\tilde{y}^s_{ij}$, as a solution of the differential equation,
\begin{align}
 \frac{\rmd}{\rmd s}\tilde{y}^s_{ij}(x) = \frac{1}{\sqrt{2}}\,\hat{\sfv}^{(s,V)}_{ij}(x) \, . 
 \end{align}
It reads
\begin{align}
 \tilde{y}^s_{ij}(x) = \tilde{y}_{ij} + \frac{1}{\sqrt{2}}\,\sum_{n=1}^\infty \frac{s^n}{n!} \bigl(\Lie_v^{n-1}\hat{\sfv}\bigr)_{ij}(x) 
 = \tilde{y}_{ij} + \frac{1}{\sqrt{2}}\,\zeta^{(s,V)}_{ij}(x) \,.
\end{align}
This is again defined only up to the equivalence relation, $\tilde{y}_{ij}\simeq \tilde{y}_{ij}+\partial_{[i}\alpha_{j]}$. 
Using this dual coordinates, we can again confirm the composition law property as in section \ref{sec:composition}. 

\section{Discussions}
\label{sec:discussion}
In this paper, we proposed a new approach for finite transformation in DFT and EFT using a geometric quantity that describes the local embedding of a null surface $\cN$ (on which the physical fields are defined) inside doubled or exceptional spacetime. 
Our approach transforms not only conventional coordinates but also dual coordinates, similar to Hohm and Zwiebach's approach. 
However, ours is free from  Papadopoulos problem the latter approach is afflicted by. 
Our approach put the condition $\tilde{\partial}^m=0$ at the outset, similar to Hull's approach and different from Hohm and Zwiebach's approach, thus treating the conventional coordinates $x^m$ and the dual coordinates $\tilde{x}_m$ on different footing. 
Our approach has the advantage that the composition of finite transformations is satisfied manifestly covariant. 
Our approach has another advantage that it can be straightforwardly extended to the SL(5) EFT to its finite transformations. 

Our approach is easily applicable to other EFTs once we have the twist matrix $L^M{}_N$ that defines the untwisted vector. 
In EFT, a parameterization of the generalized metric $\cM_{MN}$ of the form $\cM_{MN} = L^K{}_M\,L^L{}_N\,\hat{\cM}_{KL}$ is found in \cite{Berman:2011jh}, where the twist matrix, $L^M{}_N$, is composed only of the gauge potentials while the untwisted generalized metric, $\widehat{\cM}_{KL}$, is composed only of the metric. 
Using the twist matrix, we can obtain the finite transformation law in EFT for $E_n$ for $n\leq 7$ in the M-theory section (see \cite{Godazgar:2013rja} for $E_8$). 
In SL(5) EFT, by choosing a suitable section, we can also reproduce the type IIB supergravity \cite{Blair:2013gqa,Blair:2014zba}. 
In this case, the twisting is given by the type IIB gauge potentials and the resulting finite transformation law will have a different form. 
In addition, we can also consider a non-geometric twisting using the $\Omega^{ijk}$ field \cite{Malek:2012pw,Blair:2014zba}. 
The non-geometric twisting is necessary, for example, in describing the background of the exotic $5^3$-brane (see \cite{deBoer:2012ma,Blair:2014zba}). 
In the case of DFT, since the number of the dual directions was equal to the that of the original directions, the difference between the $H$-twisting and the $R$-twisting was just in the position of the indices. 
In the case of SL(5) EFT, as the two numbers differ each other, the situation will be slightly changed. 
In general EFTs up to $E_7$, the parameterization in the IIB section and non-geometric parameterization in the M-theory/IIB section will be identified explicitly in our ongoing work \cite{ours}, and the investigation of the finite transformation laws in such cases will be relegated to our separate future works. 

In this paper, we considered only a finite coordinate transformation that is connected to the identity. 
It may be possible to apply our approach also to various semi-direct products of connected / disconnected diffeomorphisms in $d$-dimensions and a finite $B$-transformation connected / disconnected to the identity, with rich topological classifications. 
It would be also interesting to study the consistency of our finite transformation laws from the viewpoints of the string worldsheet theory or the membrane worldvolume theory. 
In the duality covariant formulation of such theories such as the double sigma model, the string worldsheet is embedded into the doubled spacetime. Under a generalized diffeomorphism, the string worldsheet should be mapped according to our coordinate transformation, $x^M\to x_s^M$, up to the trivial coordinate gauge symmetry generated by a trivial Killing vector.  The consistency check, such as whether the equations of motion in the worldsheet theory is covariant under the generalized diffeomorphism, will be an important task to be confirmed. 

\section*{Acknowledgments}
We wishes to thank Martin Cederwall, Chris Hull and George Papadopoulos for useful discussions. We are grateful to Junyeong Ahn for initial discussions on the topic of this paper, and to Kanghoon Lee and Hisayoshi Muraki for helpful discussions. We acknowledge organizers and participants of the CERN-CKC TH Institute `Duality Symmetries in String and M-Theories', APCPT Focus Programs "Liouville, Integrability and Branes (11)" and "Duaity and Novel Geometry in M-Theory" for stimulating environment during this work. This work was supported in part by the National Research Foundation Grants 2005-0093843, 2010-220-C00003 and 2012K2A1A9055280.

\end{document}